\documentclass[twocolumn]{aastex62}

\newcommand{\Feii}{Fe~{\sc ii}}
\newcommand{\kms}{km\,s$^{-1}$}
\newcommand{\cmc}{cm$^{-3}$}
\newcommand{\cms}{cm$^{-2}$}

\received{...}
\revised{...}
\accepted{...}
\submitjournal{ApJ}

\shorttitle{Dense cores, filaments and outflows in S255IR}
\shortauthors{Zinchenko et al.}

\begin{document}

\title{Dense Cores, Filaments and Outflows in the S255IR Region of High Mass Star Formation}

\correspondingauthor{Igor Zinchenko}
\email{zin@appl.sci-nnov.ru}

\author[0000-0003-2793-8229]{Igor I. Zinchenko}
\affil{Institute of Applied Physics of the Russian Academy of Sciences \\
46 Ul'yanov~str., 603950 Nizhny Novgorod, Russia 
}

\author{Sheng-Yuan Liu}
\affiliation{Institute of Astronomy and Astrophysics, Academia Sinica \\
P.O. Box 23-141, Taipei 10617, Taiwan, R.O.C.
}

\author{Yu-Nung Su}
\affiliation{Institute of Astronomy and Astrophysics, Academia Sinica \\
P.O. Box 23-141, Taipei 10617, Taiwan, R.O.C.
}

\author{Kuo-Song Wang}
\affiliation{Institute of Astronomy and Astrophysics, Academia Sinica \\
P.O. Box 23-141, Taipei 10617, Taiwan, R.O.C.
}

\author{Yuan Wang}
\affiliation{Max-Planck-Institut f\"ur Astronomie \\
K\"onigstuhl 17, D-69117 Heidelberg, Germany}



\begin{abstract}

We investigate at a high angular resolution the spatial and kinematic structure of the \object{S255IR} high mass star-forming region, which demonstrated recently the first disk-mediated accretion burst in the massive young stellar object.
The observations were performed with ALMA in Band 7 at an angular resolution $ \sim 0{\farcs}1 $, which corresponds to $ \sim 180 $~AU. 
The 0.9~mm continuum, C$^{34}$S(7--6) and CCH $N=4-3$ data show a presence of very narrow ($ \sim 1000 $~AU), very dense ($n\sim 10^7$~\cmc) and warm filamentary structures in this area. At least some of them represent apparently dense walls around the high velocity molecular outflow with a wide opening angle from the S255IR-SMA1 core, which is associated with the NIRS3 YSO. This wide-angle outflow surrounds a narrow jet. At the ends of the molecular outflow there are shocks, traced in the SiO(8--7) emission. The SiO abundance there is enhanced by at least 3 orders of magnitude. 
The CO(3--2) and SiO(8--7) data show a collimated and extended high velocity outflow from another dense core in this area, SMA2. {The outflow is bent and consists of a chain of knots, which may indicate periodic ejections possibly arising from a binary system consisting of low or intermediate mass protostars. The C$^{34}$S emission shows evidence of rotation of the parent core.}
{Finally, we detected two new low mass compact cores in this area (designated as SMM1 and SMM2)}, which may represent prestellar objects.

\end{abstract}



\section{Introduction} \label{sec:intro}
The process of massive star formation attracts an enhanced attention nowadays. {This is caused by serious problems in theoretical understanding of this phenomenon and difficulties in its observational studies, which are related to the rarity and large distances of high-mass star-forming (HMSF) regions. There is a number of recent works on this subject, both observational and theoretical. Most of the results are summarized in several reviews \citep[e.g.][]{McKee07, Zinnecker07, Krumholz09, Krumholz12, Tan14, Motte18}.} A crucial question is whether {this process} can be considered as a scaled-up version of the low mass star formation or not. In this respect observations of the phenomena related to the disk accretion scenario in {HMSF regions} are very important. 

Recently this scenario got a strong support by observations of the first disk-mediated accretion burst in the massive ($ \sim 20$~M$_\odot $) young stellar object \object{S255 NIRS3} \citep{Caratti16}, accompanied by the 6.7~GHz methanol maser flare \citep{Moscadelli17, Szymczak18}. {The bolometric luminosity increased by a factor of $ \sim 5.5 $ \citep{Caratti16}. A detailed IR light curve was obtained by \cite{Uchiyama19}.} The S255IR clump has been a target for many studies over the last years. The distance to the source derived from the annual parallax measurements of water masers is $ 1.78^{+0.12}_{-0.11} $~kpc \citep{Burns16}. Observations with the SMA revealed three major cores within this clump \citep{Wang11,Zin12}. The NIRS3 source is associated with one of them, S255IR-SMA1, which represents a disk around the massive young stellar object \citep{Zin15}. It is associated with an ionized jet \citep{Howard97,Cesaroni18}. Our ALMA data show a submillimeter burst in this object, {which lasted about 2 years} \citep{Liu18}. {An even more impressive} burst {(with an increase by a factor of $ \sim 70 $ in luminosity but with a lower initial level)} was detected in {another} massive object NGC6334I-MM1 \citep{Hunter17}. {These events demonstrate an importance of the episodic disk accretion in the process of high mass star formation. They are consistent with theoretical models, which predict disk fragmentation and episodic accretion of the fragments onto a massive protostar \citep{Meyer17, Meyer19}.}

{Previous single-dish and interferometric studies have shown an extended outflow from S255IR, which was attributed to the SMA1 as a driving source \citep{Wang11, Zin15}. However our ALMA data reveal almost parallel bipolar outflows from both SMA1 and SMA2 cores \citep{Zin18-raa}. The outflow from SMA1 has a wide opening angle. At the same time, as mentioned above, there is a narrow ionized jet from SMA1. This suggests a two-wind outflow, which is actually expected in theoretical models \citep[e.g.][]{Arce07, McKee07}. However observations of such outflows in HMSF regions are very limited \citep{Anglada18}.}

An interesting feature found in our ALMA observations of S255IR is an unusual very narrow and apparently very dense filament \citep{Zin18-raa,Zin17-iau-s255ir}.
Interstellar filaments are now the subject of numerous investigations, since they are considered as primary sites of star formation \citep[e.g.][]{Andre14,Li16}. 

{This paper is devoted to the surroundings of this remarkable object S255IR, which are apparently strongly influenced by its activity.}
We present additional data on this region and discuss its morphology, kinematics and physical properties {with an emphasis on the new and unusual features mentioned above}.

\section{Observations}
The observations were performed with ALMA in Band 7 during several observing sessions in 2016--2017 under the project 2015.1.00500.S. Details of the observations in 2016 are given in \cite{Zin17}. Briefly, four spectral windows were observed, centered at around 335.4 GHz, 337.3 GHz, 349.0 GHz, and 346.6 GHz, with bandwidths of 1875.0 MHz, 234.4 MHz, 937.5 MHz, and 1875.0 MHz.
Details of the observations in 2017 are given in \cite{Liu18}. The same spectral windows were observed. 
The projected array baselines for all sessions range between 15~m and 3.0~km. 

{Here we discuss the continuum emission and the CO $ J=3-2 $ line at 345.8~GHz, C$^{34}$S $ J=7-6 $ line at 337.4~GHz, SiO $ J=8-7 $ line at 347.3~GHz and CCH $ N=4-3 $ lines at 349.4~GHz. The channel width was 244~kHz for C$^{34}$S and CCH (0.22 and 0.21~\kms, respectively), and 977~kHz for CO and SiO (0.85 and 0.84~\kms, respectively). The spectral resolution was twice of these values with Hanning-smoothing applied.  
The resulting images achieve an angular resolution of $ 0{\farcs}11\times 0{\farcs}14 $ (PA = --6.9$^\circ$) for C$^{34}$S, $ 0{\farcs}11\times 0{\farcs}14 $ (PA = --13.0$^\circ$) for CCH, $ 0{\farcs}10\times 0{\farcs}15 $ (PA = --5.3$^\circ$) for CO, $ 0{\farcs}11\times 0{\farcs}16 $ (PA = 4.4$^\circ$) for SiO and $ 0{\farcs}11\times 0{\farcs}15 $ (PA = --3.2$^\circ$) for continuum with Briggs weighting with a robust parameter of 0.5.}

\section{Results}
{We present the measurement results in the flux density units and/or in the Rayleigh-Jeans (RJ) temperature scale, which is just a linear measure of intensity. It is frequently defined as the \textit{radiation temperature} ($ T_R $) and is related to the source parameters by the simple expression, which follows from the equation of radiative transfer \citep[e.g.][]{Mangum15}}:
\begin{equation} \label{eq:tr}
T_R = f\left[ J_\nu(T) - J_\nu(T_{bg})\right] \left[ 1-\exp{\left(-\tau_\nu \right)}\right] \;,
\end{equation}
where
\begin{equation}
J_\nu(T) \equiv \frac{\frac{h\nu}{k}}{\exp{\left(\frac{h\nu}{kT}\right)}-1} \;,
\end{equation}
{$h$ is the Planck constant, $k$ is the Boltzmann constant, $f$ is the so-called beam filling factor, $\tau_\nu$ is the source optical depth. The temperature $T$ here can be the excitation temperature of the considered molecular transition or the dust temperature. It is assumed to be constant along the line of sight. $T_{bg}$ is the background temperature, which usually refers to the temperature of the cosmic microwave background radiation.}

\begin{figure*}
\begin{minipage}[b]{0.68\textwidth}
\includegraphics[width=\columnwidth]{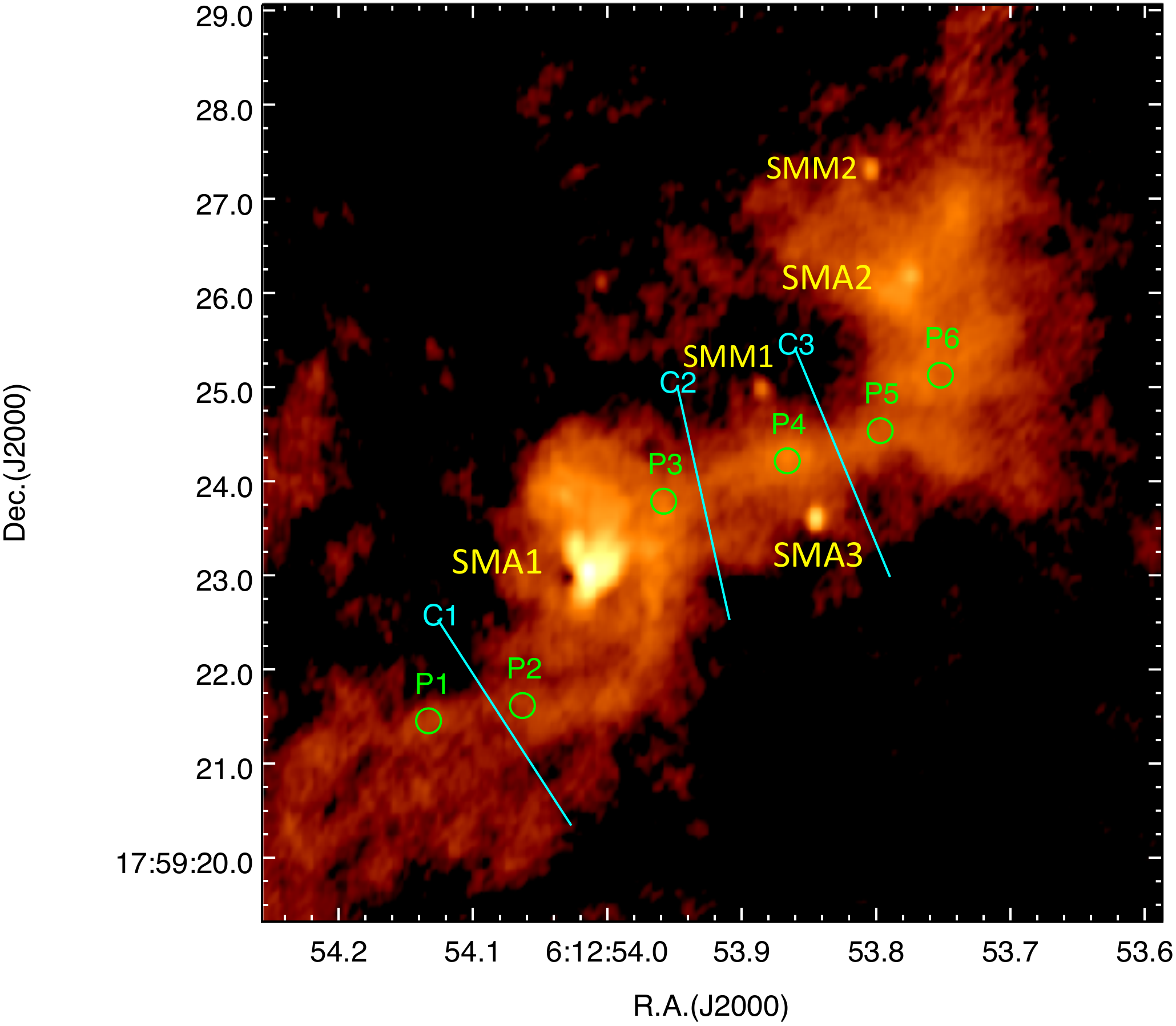} 
\includegraphics[width=\columnwidth]{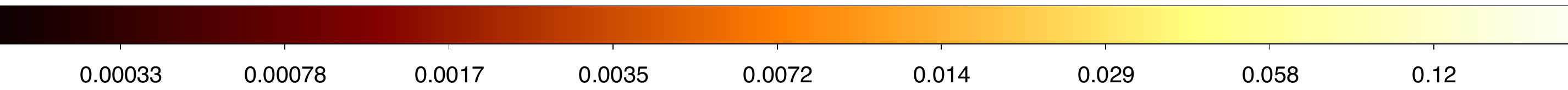} 
\end{minipage}
\hfill
\begin{minipage}[b]{0.31\textwidth}
\includegraphics[width=\columnwidth]{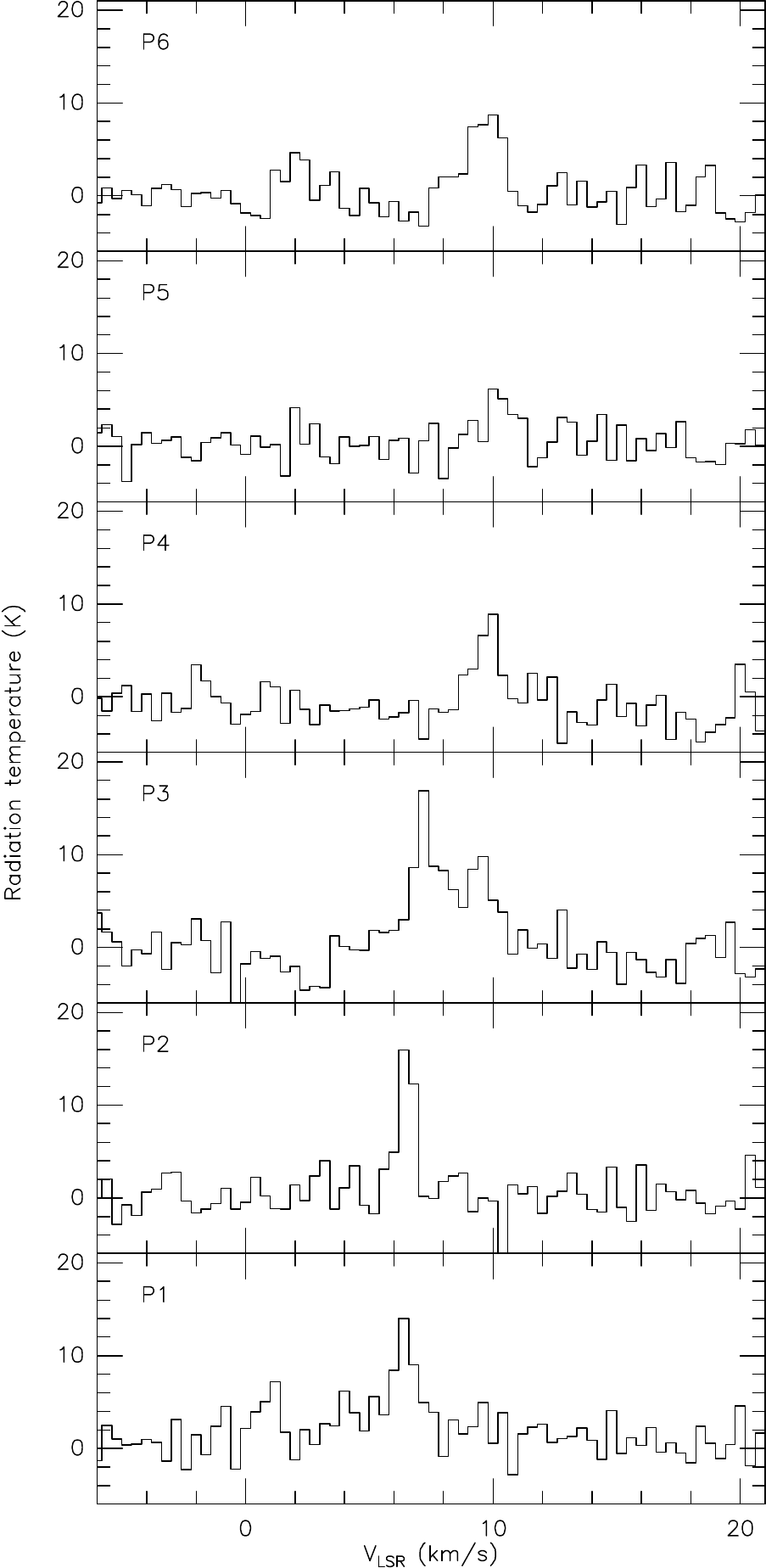} 
\end{minipage}
\caption{Left panel: The image of the S255IR region in continuum at 0.9~mm (logarithmic scale). {The brightness scale in Jy/beam is indicated by the colorbar.} Several positions (P1...P6) are indicated by the green circles and three cuts (C1, C2, C3) by the cyan lines. Right panel: The C$^{34}$S $J=7-6$ spectra at the selected positions indicated in the left panel.}
\label{fig:filament}
\end{figure*}

The ALMA image of the S255IR region in continuum at 0.9~mm is shown in Fig.~\ref{fig:filament}. A filamentary structure can be easily seen. Several compact sources are visible, too. Among them there are three earlier identified objects, labelled as SMA1, SMA2 and SMA3 \citep{Wang11,Zin12}. The SMA1 and SMA2 clumps seem to be located within this structure, while the SMA3 clump looks isolated. In addition there are two more point-like sources, which we label as SMM1 and SMM2. The estimates of the fluxes and sizes of the continuum sources obtained from 2D Gaussian fit are given in Table~\ref{table:cont-mm}. These fluxes and sizes, especially for the SMA2, are somewhat uncertain because the sources are observed on a significant background, which has a complicated morphology. We do not include in the table the SMA1 clump since its flux is variable \citep{Liu18} and the brightness distribution is apparently non-Gaussian. In the presented image its flux integrated in the circle of {0.5 arcsec in radius centered at R.A.(J2000)=06$^h$12$^{m}$54.008$^s$, Dec.(J2000)=17$^\circ$59$^\prime$23$\farcs$12} is approximately 1.04~Jy. {This circle covers the brightest part of the SMA1 emission, although its size is of course still rather arbitrary.} The continuum brightness along the filamentary structure, away from the major clumps, is $\sim 2-5$~mJy/beam, which corresponds to $T_R\sim 1.5-3$~K.

\begin{deluxetable}{lllcccr}
\tablecaption{Names, positions, flux densities, deconvolved angular sizes and position angles of the submillimeter wave continuum sources measured at 0.9~mm. \label{table:cont-mm}}
\tablehead{\colhead{Name} &\colhead{$\alpha$(2000)} &\colhead{$\delta$(2000)} &\colhead{$S_{0.9}$} &\colhead{$\theta_{\mathrm{max}}$} &\colhead{$\theta_{\mathrm{min}}$} &\colhead{P.A.} \\ \colhead{} &\colhead{(h m s)} &\colhead{($^\circ$ $^\prime$ $^{\prime\prime}$)} &\colhead{(mJy)} &\colhead{($^{\prime\prime}$)} &\colhead{($^{\prime\prime}$)} &\colhead{($^\circ$)} }
\tablecolumns{8}
\startdata
SMA2 &06:12:53.775 &17:59:26.17 &60  &0.37    &0.25  &157\\
SMA3 &06:12:53.843 &17.59.23.62 &48 &0.11 &0.09 &8\\
SMM1 &06:12:53.884 &17.59.24.99 &15 &0.20 &0.14 &17 \\
SMM2 &06:12:53.805 &17.59.27.30 &10 &0.15 &0.12 &3
\enddata

\end{deluxetable}

The total flux density in continuum is {approximately} 6.6~Jy. The summary flux density of the compact cores is $\approx 1.2$~Jy. The physical conditions in the vicinity of these cores should be different from those in the bulk of the filament. Therefore, for the estimation of the filament parameters we subtract the contribution of these clumps and arrive at an estimate for the filament flux density of 5.4~Jy.

\begin{figure}
\includegraphics[width=\columnwidth]{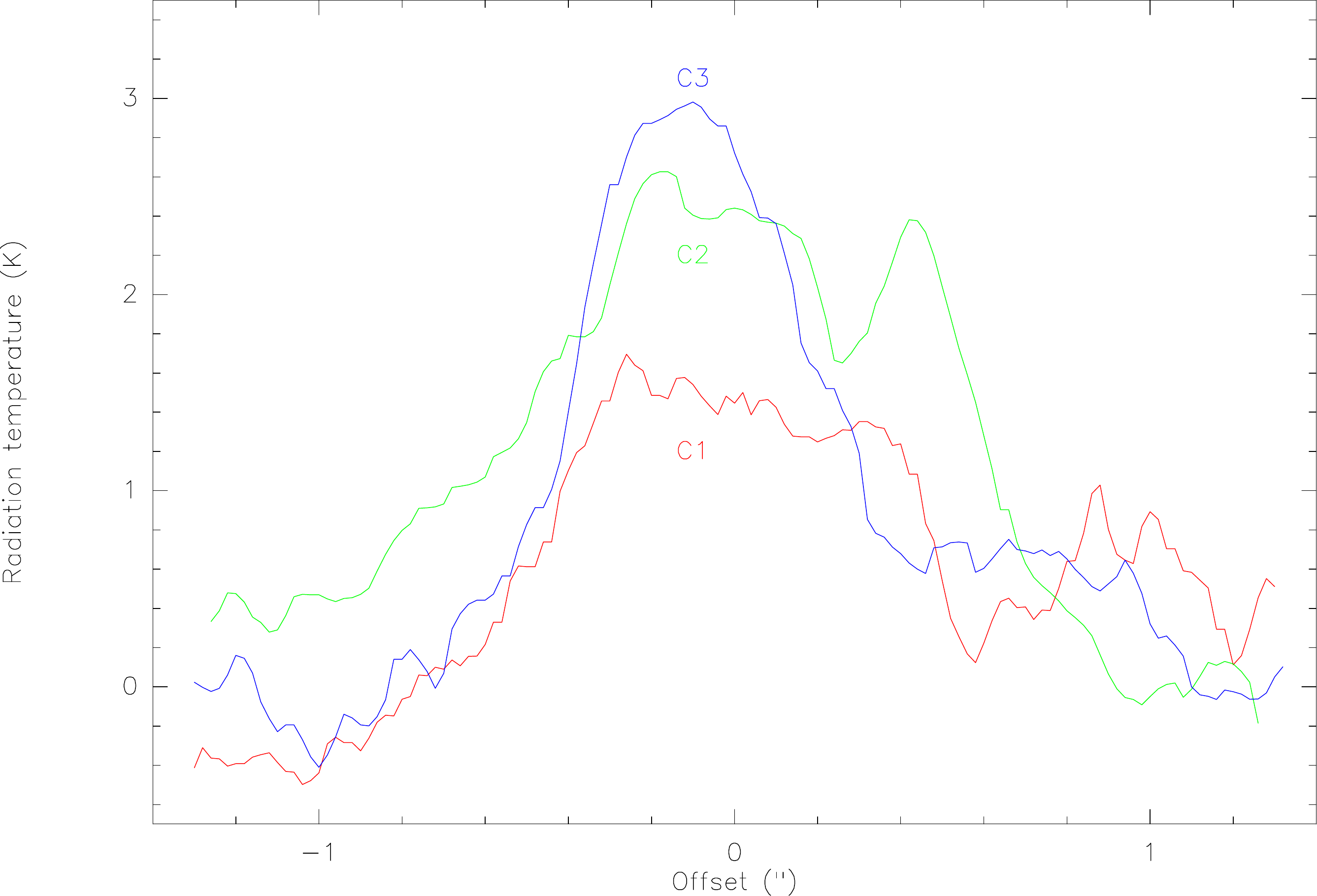} 
\caption{Continuum intensity profiles for three cuts across the filamentary structure (C1, C2 and C3), indicated in Fig.~\ref{fig:filament}.}
\label{fig:cuts}
\end{figure}

In Fig.~\ref{fig:cuts} we show the intensity profiles along the cuts indicated in Fig.~\ref{fig:filament}. It is easy to see that the filament width at the half intensity level is $\sim 0.6-1$ arcsecond, which corresponds to $\sim 1000-1800$~AU. The length of the filamentary structure {along the emission ridge} is about 15$^{\prime\prime}$, which corresponds to 27000~AU or 0.13~pc.

In the molecular line data we see the C$^{34}$S $J=7-6$ emission along the filament. The {single pixel} C$^{34}$S spectra at the selected positions indicated in Fig.~\ref{fig:filament} (P1...P6) are shown in the same figure.

One can see that the C$^{34}$S velocities in the southern and in the northern parts of the filamentary structure are different. In the south the peak velocity is about 6.5~\kms, while in the north it is about 10~\kms. Near the SMA1 clump we see both components. The line widths of the individual components {obtained from Gaussian fits taking into account instrumental broadening are from $ 0.6\pm 0.1$~\kms\ at the P2 position to $ 1.5\pm 0.2$~\kms\ at the P6 position}.

\begin{figure}
\includegraphics[width=\columnwidth]{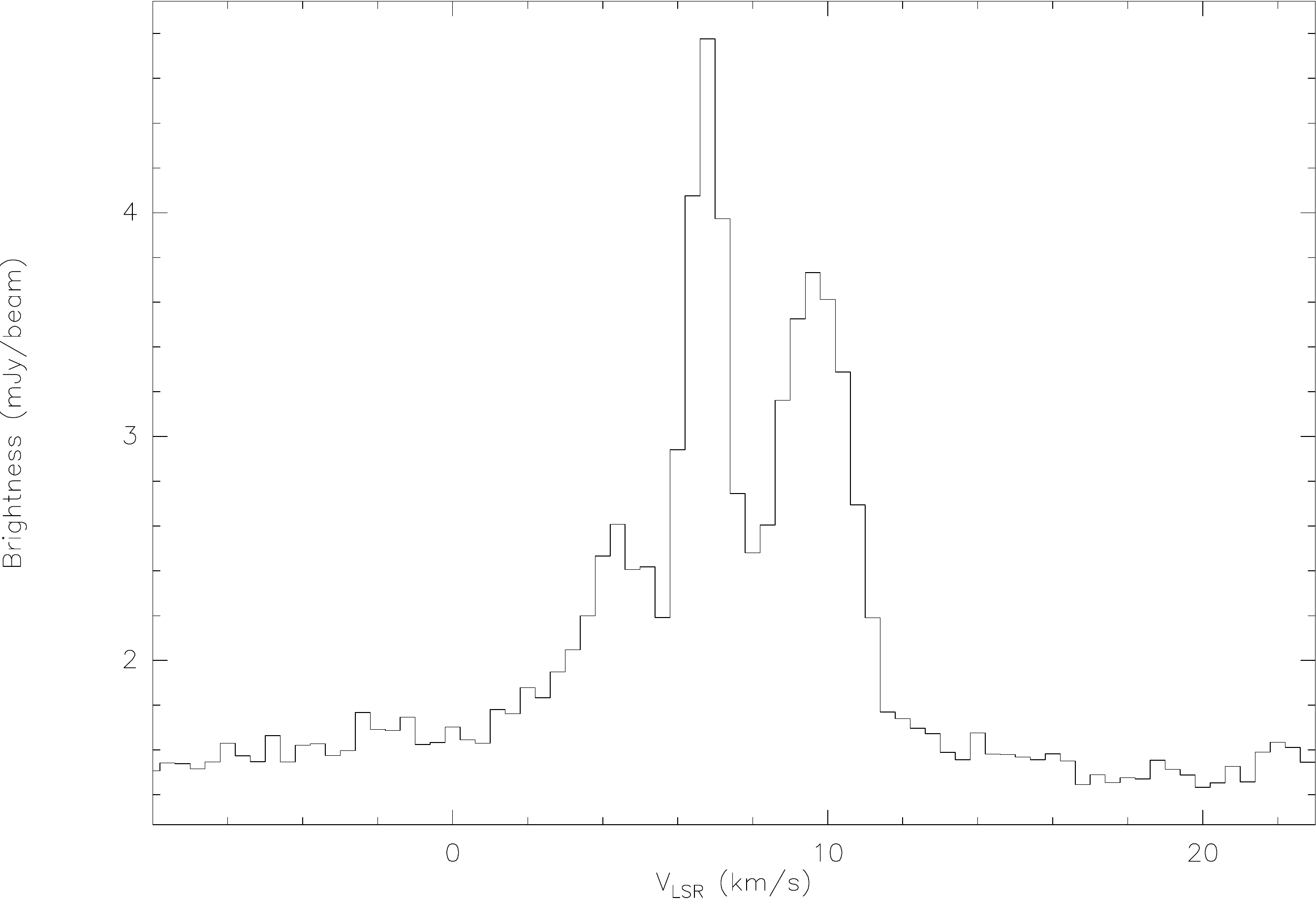} 
\caption{The average C$^{34}$S(7--6) spectrum in the S255IR area {(indicated in Fig.~\ref{fig:channel})}.}
\label{fig:c34s-aver}
\end{figure}

\begin{figure*}
\begin{minipage}[b]{0.49\textwidth}
\includegraphics[width=\textwidth]{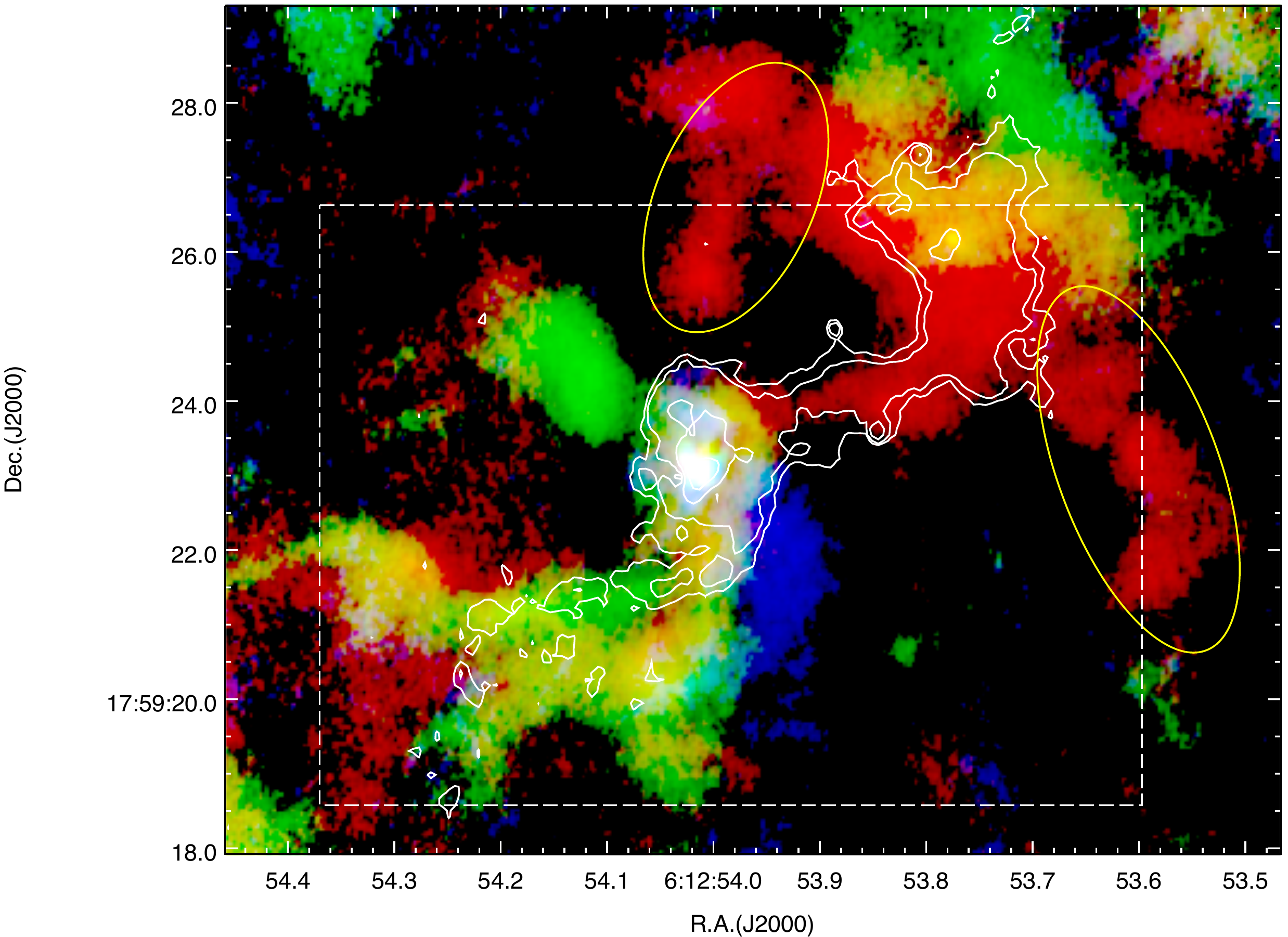}
\end{minipage}
\hfill
\begin{minipage}[b]{0.49\textwidth}
\includegraphics[width=\textwidth]{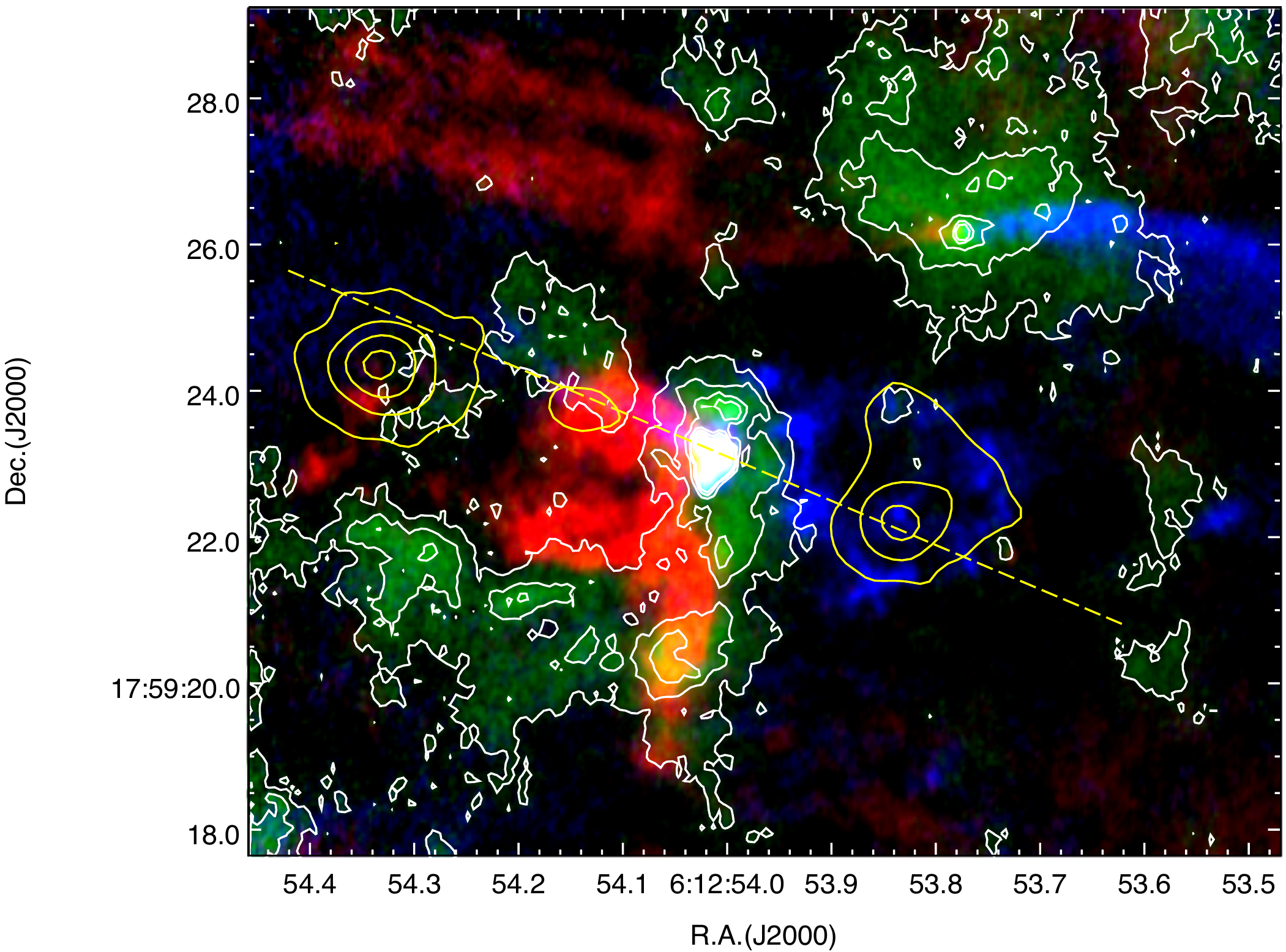}
\end{minipage}
    \caption{Left panel: a 3-color map of the C$^{34}$S $J=7-6$ emission in the S255IR area. The emission in the velocity intervals 9--11~\kms, 5.5--7.5~\kms and 3--5~\kms\ is shown in red, green and blue colors, respectively. The white contours show the continuum emission at 0.9~mm as in Fig.~\ref{fig:filament}. The contour levels are at 2, 3, 9, 42.5 and 230~mJy\,beam$^{-1}$. The dashed rectangle shows the area used to compute the average C$^{34}$S spectrum in Fig.~\ref{fig:c34s-aver}. The ellipses indicate the arc-like structures discussed in the text. Right panel: The red and blue colors show the high-velocity red-shifted {(integrated in the velocity range from 20 to 90~\kms)} and blue-shifted {(integrated in the velocity range from --67 to --16~\kms)} CO(3--2) emission, respectively. {The C$^{34}$S emission integrated in the velocity range 1--13~\kms\ is shown in green and by the white contours. The contour levels are from 20 to 200~mJy\,beam$^{-1}$\,\kms\ in step of 45~mJy\,beam$^{-1}$\,\kms.} The yellow contours show the \Feii\ emission \citep{Wang11}. The dashed line indicates the jet axis as found by \cite{Howard97} {from near-IR H$_2$ and Br$\gamma$ images}.}
    \label{fig:channel}
\end{figure*}

A more detailed inspection of the data shows that the C$^{34}$S(7--6) emission is observed in a rather large area around the SMA1--SMA3 clumps. In the average C$^{34}$S(7--6) spectrum three main velocity components can be distinguished (Fig.~\ref{fig:c34s-aver}). In addition to the two components mentioned above, there is an emission peak at $\sim 4$~\kms. A general picture of the C$^{34}$S emission in this area is presented in Fig.~\ref{fig:channel}. The three main velocity components, at 4~\kms, 6.5~\kms\ and 10~\kms, are shown by the blue, green and red colors, respectively. It is worth mentioning that the systemic velocity of the SMA1 clump is about 5~\kms\ \citep{Zin15}. 

\begin{figure*}
\centering
\includegraphics[width=0.9\textwidth]{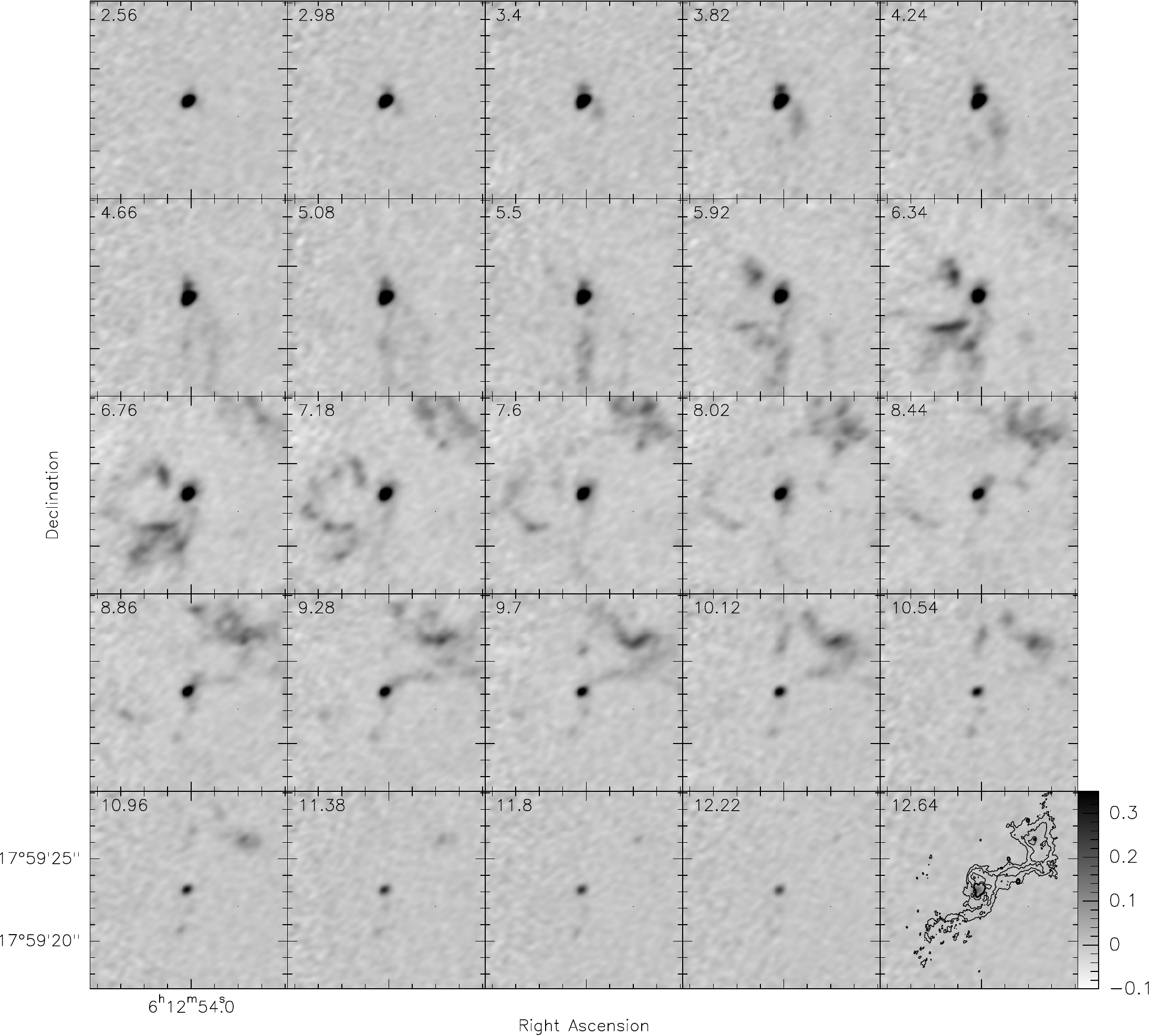} 
\caption{The channel maps of the C$^{34}$S(7--6) emission in the S255IR area. The center velocities are indicated in the upper left corner of each panel. {The channel width is 0.42~\kms. The tick spacing at the Right ascension axis is 0.1$^s$.} The color bar indicates the intensity scale (in Jy\,beam$^{-1}$). The contours in the last panel show the 0.9~mm continuum emission. {The contour levels are at 2, 4, 8, 12 and 16~mJy\,beam$^{-1}$.} }
\label{fig:c34s-chmaps}
\end{figure*}

Fig.~\ref{fig:channel} shows that distributions of the dust continuum and C$^{34}$S emission are significantly different. Some regions of a rather strong C$^{34}$S emission have no noticeable counterpart in continuum. A more detailed picture of the spatial and kinematic distribution of the C$^{34}$S emission is presented in the channel maps (Fig.~\ref{fig:c34s-chmaps}).

\begin{figure}
\includegraphics[width=\columnwidth]{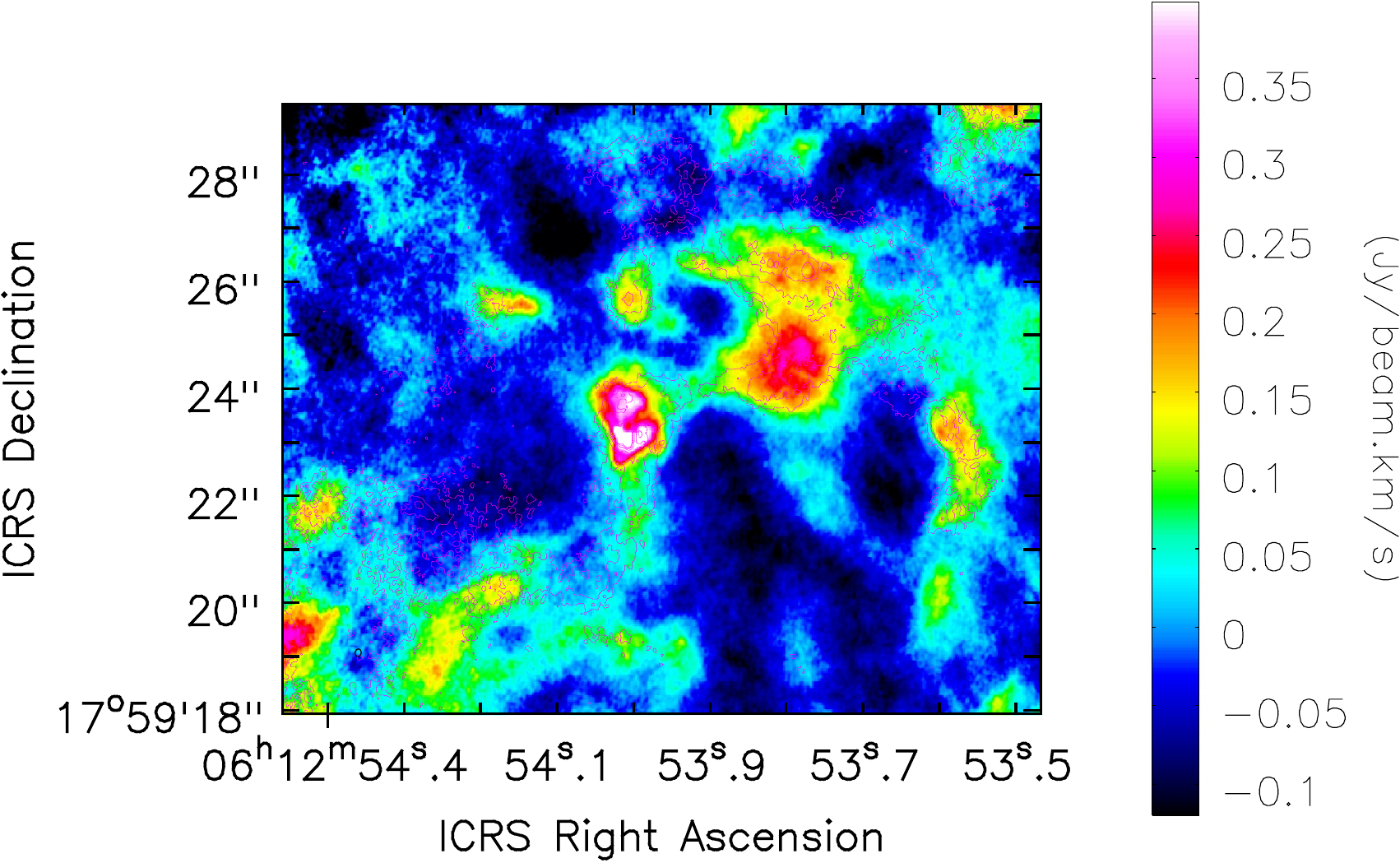} 
\caption{The image of the CCH $N=4-3$ integrated emission overlaid with contours of the C$^{34}$S $J=7-6$ emission in the 10~\kms\ component. {The contour levels are at 10, 20 to 100 in step of 20~mJy\,beam$^{-1}$\,\kms.} Around the SMA1 core the CCH lines are blended with other strong lines.}
\label{fig:cch}
\end{figure}

A further inspection of the ALMA data shows a rather strong emission in the CCH lines in this area. Four strongest components of the $N=4-3$ transition are observed: $J=9/2-7/2$, $F=5-4$; $J=9/2-7/2$, $F=4-3$; $J=7/2-5/2$, $F=4-3$ and $J=7/2-5/2$, $F=3-2$ with the excitation energies of the upper levels of about 41.9~K. In some parts the weaker $J=7/2-5/2$, $F=3-3$ component is also seen. The emission 
is {rather} well correlated spatially with the 10~\kms\ C$^{34}$S component (Fig.~\ref{fig:cch}). The {radiation} temperature in the CCH lines reaches {approximately 37~K at the emission peak}. 

{The detection of the weaker CCH line allows an estimation of the optical depth in the CCH lines under the assumption of the same excitation temperature for all transitions. The ratio of the intensities of the strongest $J=9/2-7/2$, $F=5-4$ transition and the weak $J=7/2-5/2$, $F=3-3$ transition is $ 11\pm 1$. At the same time the ratio of the line strengths of these lines is $\approx 29.2$ (CDMS -- \citealt{Mueller05,Endres16}). This gives us immediately estimates of the optical depths in the weaker transition of $\approx 0.08$ and for the strongest transition of $\approx 2.4$. From Eq.~(\ref{eq:tr}) assuming the beam filling factor equal unity we obtain the excitation temperature $T_{ex}\approx 50$~K. This gives us a lower limit for the kinetic temperature.}

{However, the assumption of the equal excitation temperatures for different transitions needs a justification. We modeled the CCH excitation with RADEX \citep{vdTak07}. This modeling shows that the observations are best reproduced at densities $ n\ga 10^7 $~cm$^{-3}$, $ T_{kin}=55 $~K and CCH column density $ N(\mathrm{CCH})\approx 5\times 10^{15} $~cm$^{-2}$. A reasonable fit can be obtained at $ n= 3\times 10^6 $~cm$^{-3}$, $ T_{kin}=60 $~K. At lower densities the excitation of different components becomes significantly different and the intensity ratios for the stronger components are inconsistent with the observations.}

{The CCH radiation temperature peaks in the area of the maximum CCH integrated intensity near the SMA2 core. Although in the other parts the radiation temperature is lower, the optical depth is also lower, as indicated by the line ratios. The estimates show that in some areas the excitation temperature can be even higher than at the emission peak (but the uncertainties of these estimates are rather large). Actually this emission peak is not associated with any heating source, so there is no apparent physical reason for an enhanced temperature at this position. Most probably the peak of the radiation temperature is explained by a higher CCH optical depth here.}

\begin{figure*}
\begin{minipage}[b]{0.49\textwidth}
\includegraphics[width=\columnwidth]{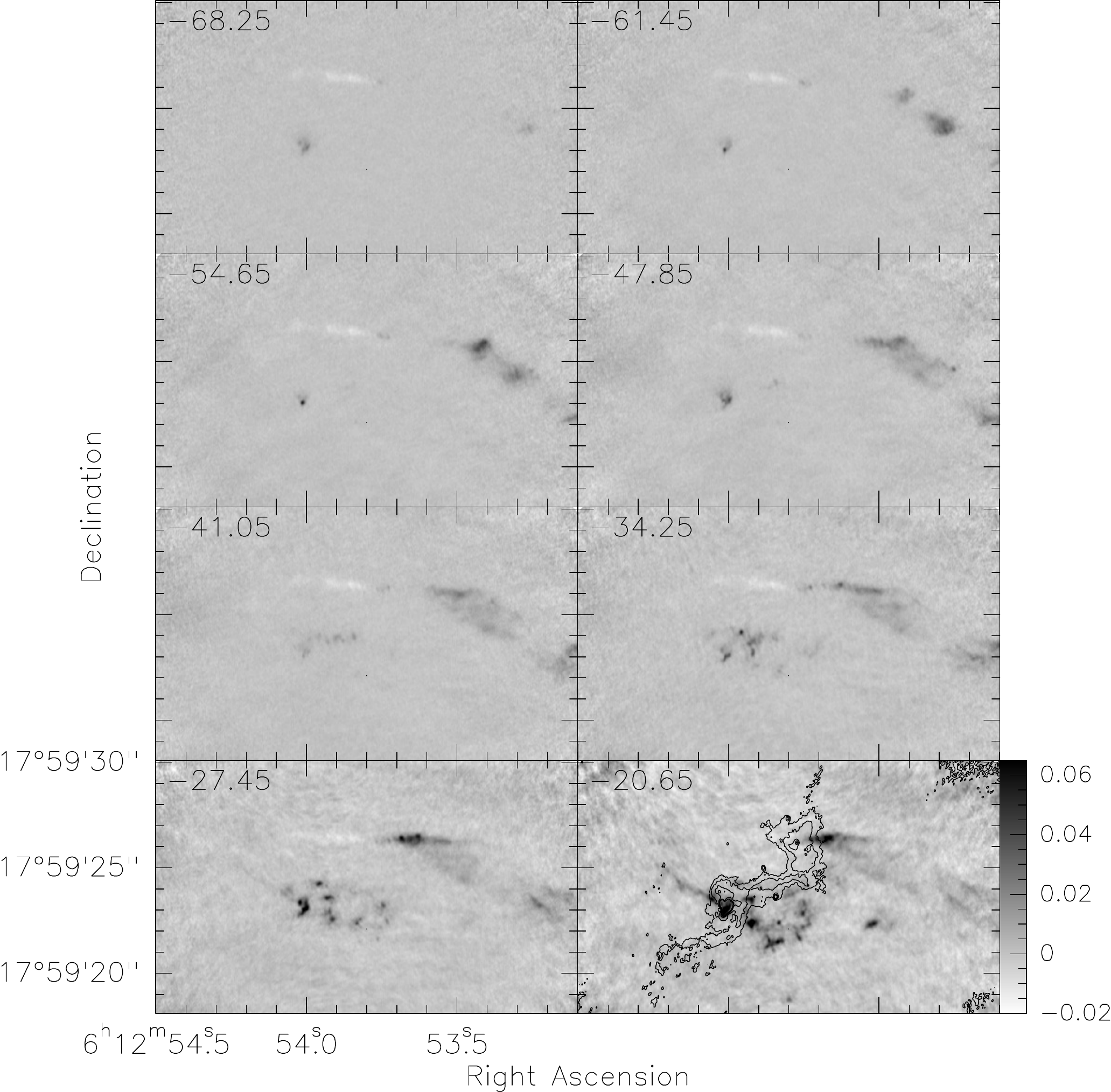} 
\end{minipage}
\hfill
\begin{minipage}[b]{0.49\textwidth}
\includegraphics[width=\columnwidth]{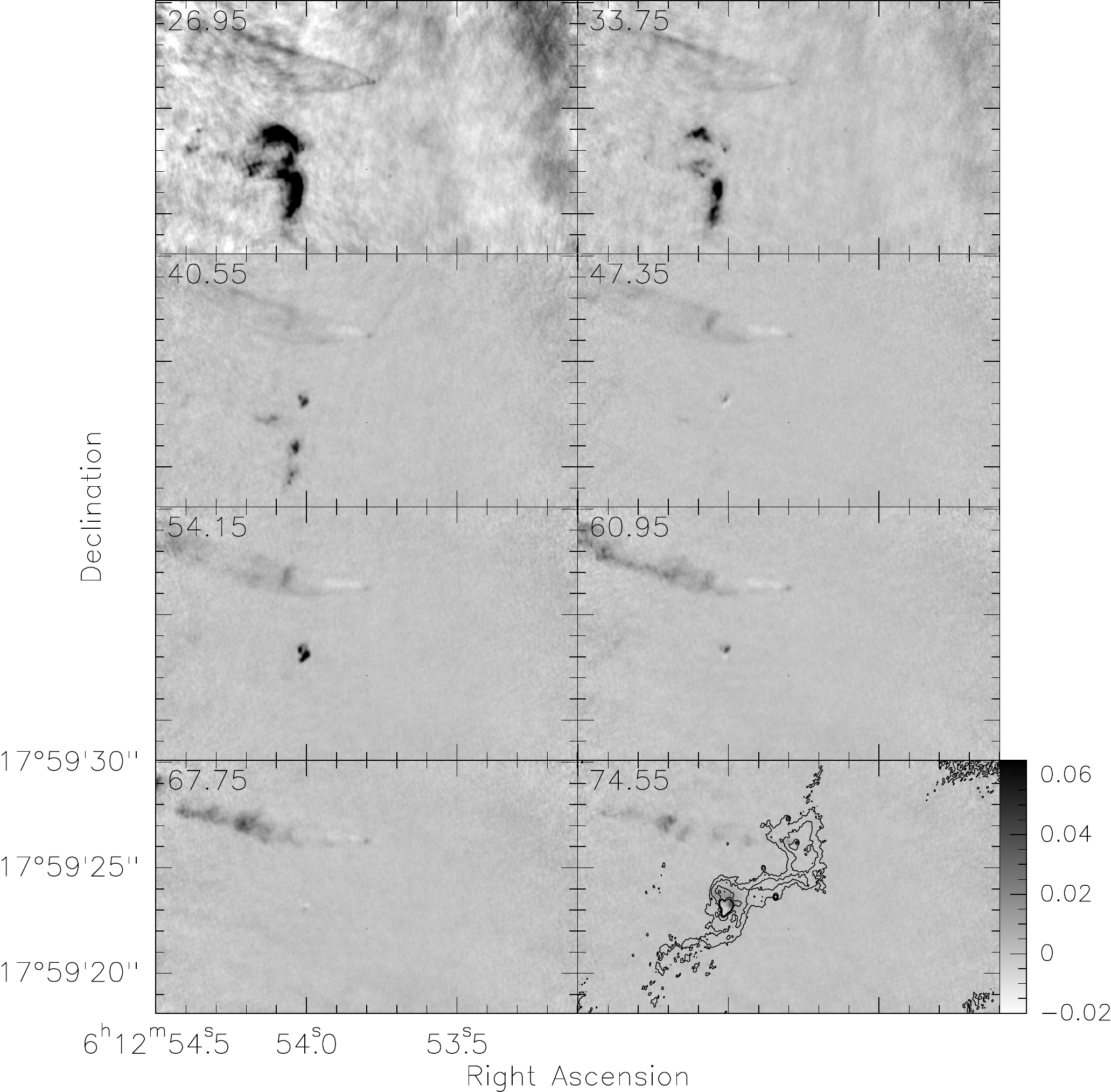} 
\end{minipage}
\caption{Channel maps of the high velocity CO(3--2) blue-shifted (left panel) and red-shifted (right panel) emission in the S255IR area. The center velocities are indicated in the upper left corner of each panel. {The channel width is 6.8~\kms.} The contours in the last panel show the 0.9~mm continuum emission. The color bar indicates the intensity scale (in Jy/beam).}
\label{fig:co-outflow}
\end{figure*}

In the right panel of Fig.~\ref{fig:channel} the high velocity CO(3--2) emission in the blue and red-shifted wings is shown in blue and red colors, respectively. A more detailed picture of the CO kinematics is given in Fig.~\ref{fig:co-outflow}, where we present channel maps of the blue-shifted and red-shifted high velocity CO(3--2) emission.

\begin{figure}
\includegraphics[width=\columnwidth]{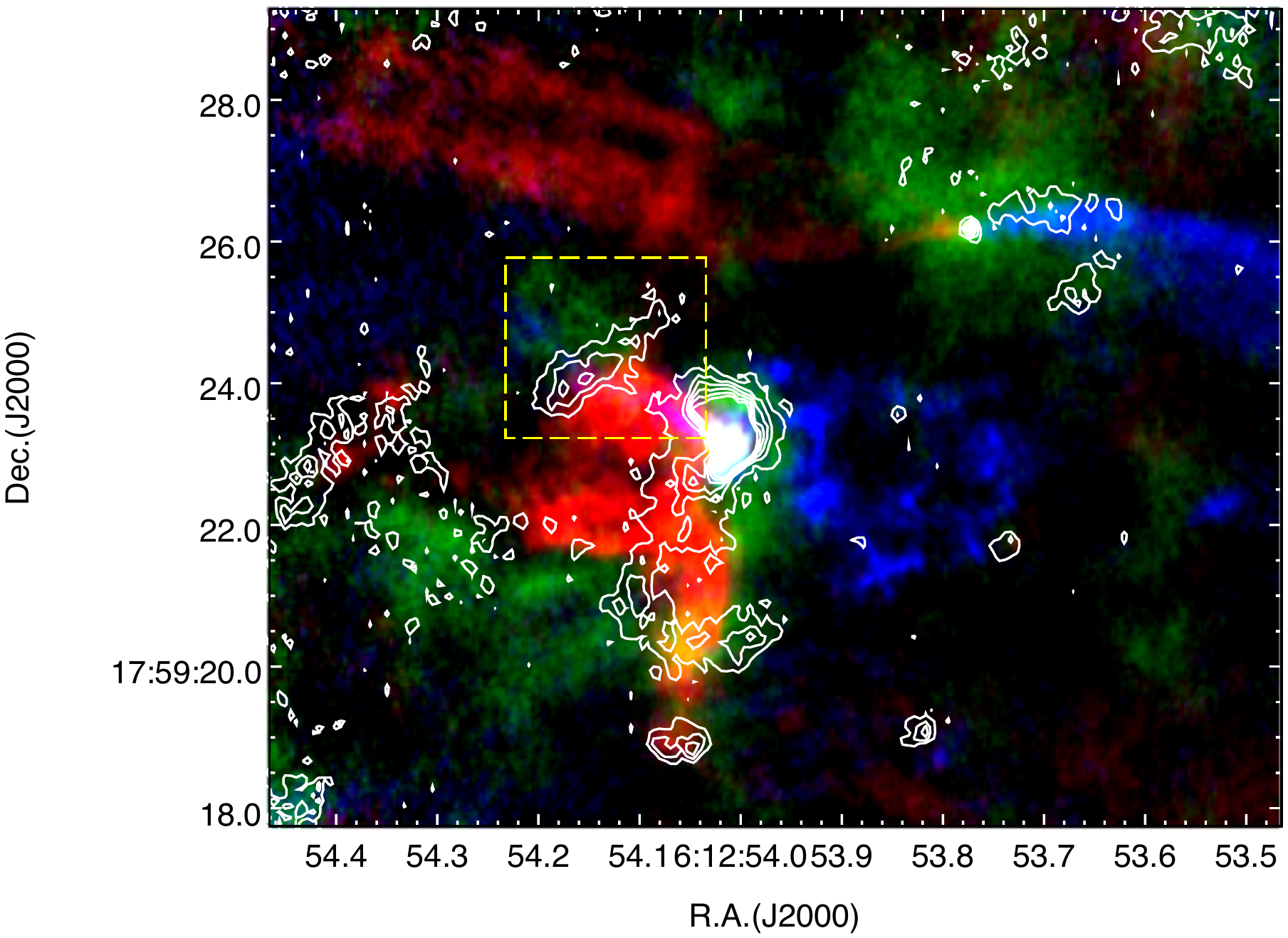} \caption{Image: the SiO(8--7) integrated emission map (white contours) overlaid on the map of the high velocity CO(3--2) emission (red and blue colors). The integrated C$^{34}$S(7--6) emission is shown in green. {The contour levels are at 75 to 450 in step of 75~mJy\,beam$^{-1}$\,\kms.}}
\label{fig:sio+cont+outflow}
\end{figure}

\begin{figure}
\includegraphics[width=\columnwidth]{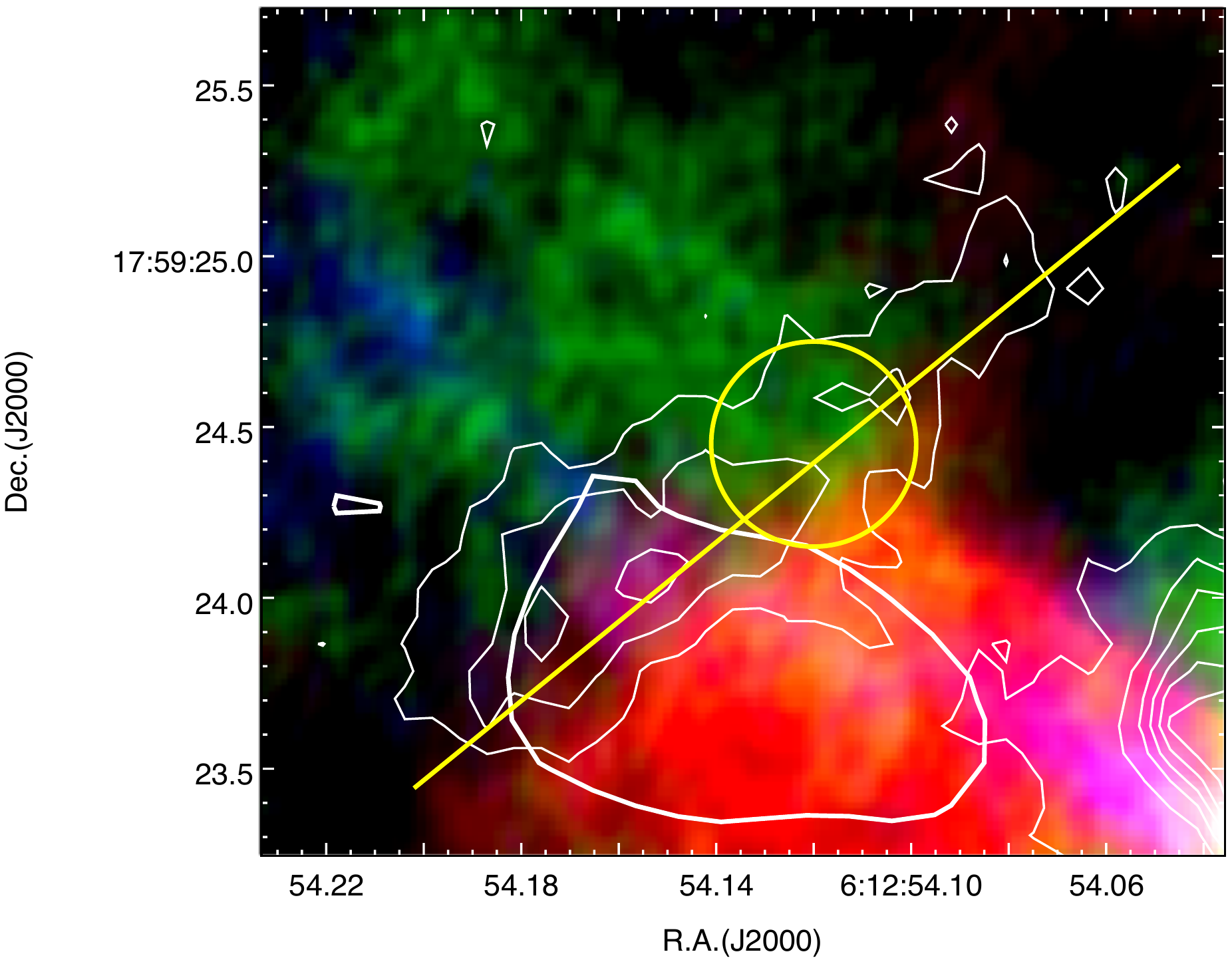} 
\caption{The close-up view of the region marked by the dashed box in Fig.~\ref{fig:sio+cont+outflow}. The thick white contours show the \Feii\ emission. The yellow circle indicates the position, where the SiO(8--7) and C$^{34}$S(7--6) spectra are presented {in Fig.~\ref{fig:sio+c34s-sp}}, and along the yellow line the position-velocity diagrams for these lines are shown in Fig.~\ref{fig:sio+c34s-pv}.}
\label{fig:sio-box}
\end{figure}

\begin{figure}
\includegraphics[width=\columnwidth]{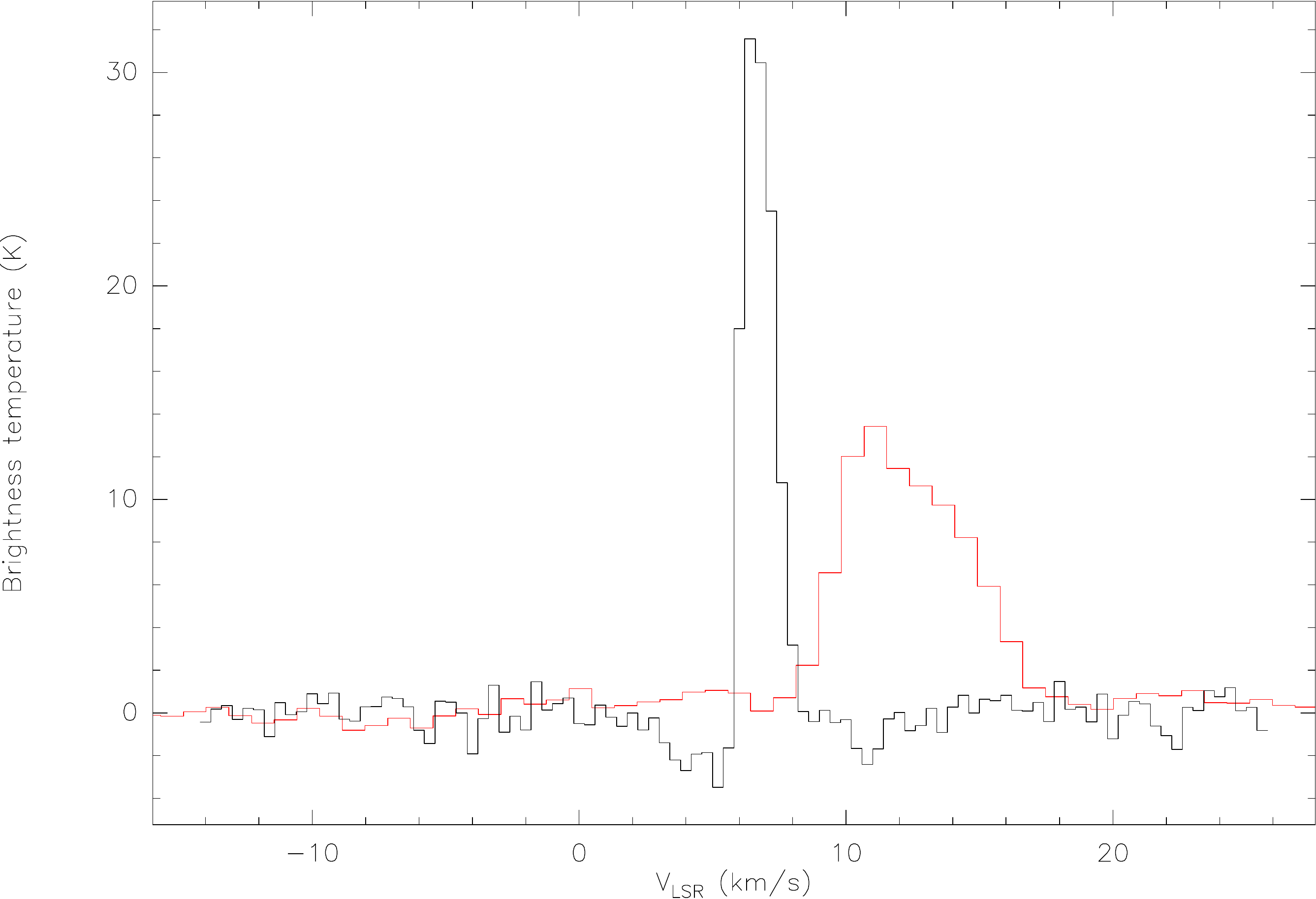} \caption{The SiO(8--7) (red line) and C$^{34}$S(7--6) (black line) spectra {averaged over} the yellow circle in Fig.~\ref{fig:sio-box}.}
\label{fig:sio+c34s-sp}
\end{figure}

\begin{figure}
\includegraphics[width=\columnwidth]{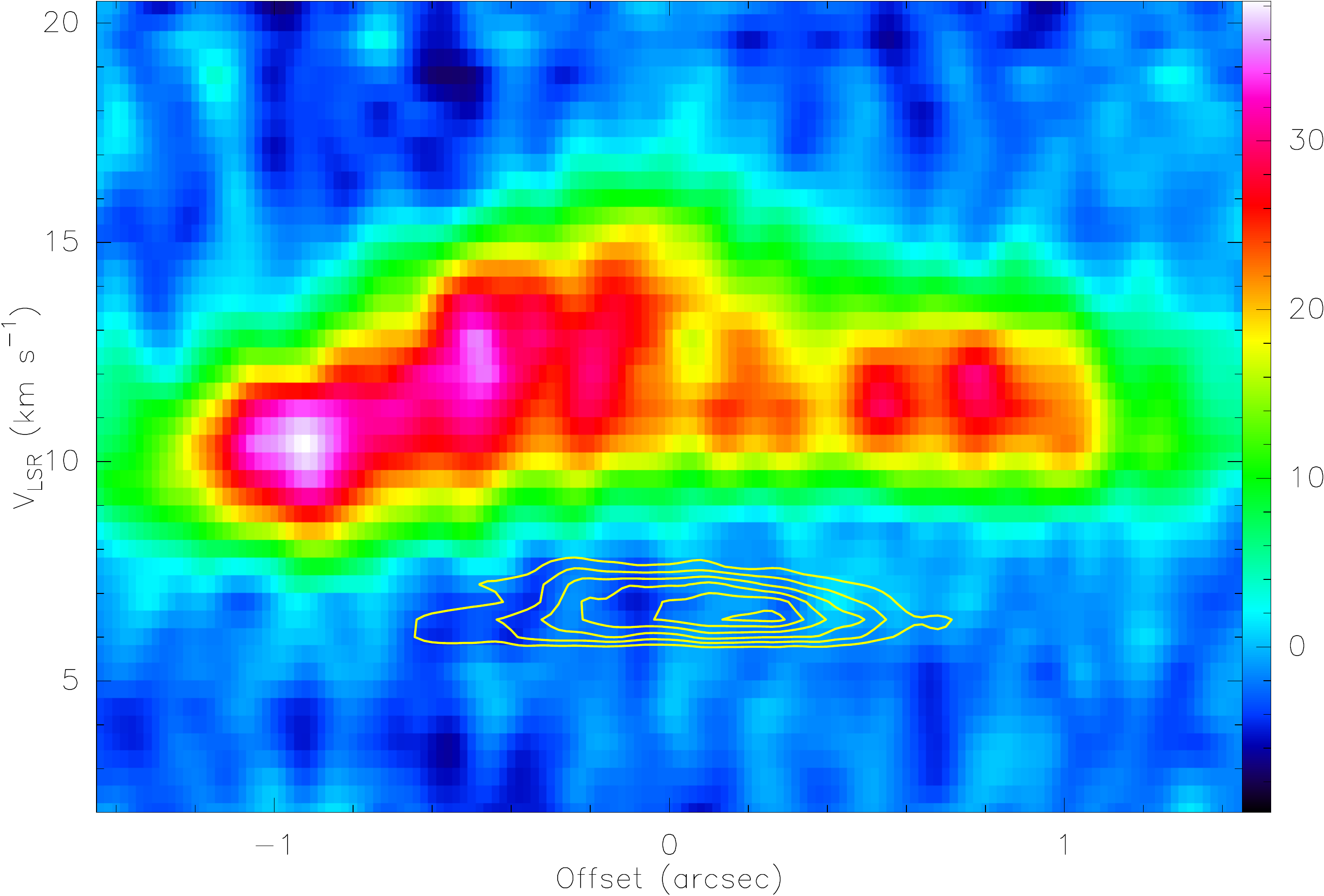} 
\caption{The position-velocity diagrams for the SiO(8--7) (image) and C$^{34}$S(7--6) (contours) along the line shown in Fig.~\ref{fig:sio-box}. {The negative offsets correspond to the SE direction. The color wedge shows the SiO intensity in mJy\,beam$^{-1}$. The contour levels are from 0.2 to 0.95 in step of 0.15 of the peak C$^{34}$S intensity, which is 58~mJy\,beam$^{-1}$.}}
\label{fig:sio+c34s-pv}
\end{figure}

In some parts of this area we also see a rather strong SiO(8--7) emission (Fig.~\ref{fig:sio+cont+outflow}). In particular it is observed toward the region outlined by the dashed box in Fig.~\ref{fig:sio+cont+outflow}. The close-up view of this region with an addition of the \Feii\ emission is presented in Fig.~\ref{fig:sio-box}. The SiO(8--7) and C$^{34}$S(7--6) spectra {averaged over the area} marked by the yellow circle in this figure are shown in Fig.~\ref{fig:sio+c34s-sp}. In Fig.~\ref{fig:sio+c34s-pv} we present the position-velocity diagrams for these lines along the path indicated in Fig.~\ref{fig:sio-box}.

\section{Discussion}
\subsection{General structure and kinematics of S255IR}
Our C$^{34}$S and CCH data reveal a rather complicated morphology and kinematics of the region. At the first glance they indicate the presence of several intersecting filamentary structures. However a comparison with the high-velocity CO(3--2) emission (Fig.~\ref{fig:channel}) {and with the SiO emission (Figs.~\ref{fig:sio+cont+outflow},\ref{fig:sio-box})} reveals another picture. One can see that {a strong C$^{34}$S emission surrounds a part of} the red-shifted lobe of the high-velocity outflow from the SMA1 core. Apparently this emission originates in the dense walls of the outflow cavity. {This view is supported by observations of the SiO emission at the interface of the CO and C$^{34}$S emission regions (Figs.~\ref{fig:sio+cont+outflow},\ref{fig:sio-box}). It is well known that an enhanced SiO emission usually traces shocked regions. Actually the morphology of the red-shifted outflow lobe is rather complicated. There is the ``main" outflow around the jet and an additional stream in the south direction, which is also associated with the SiO emission. This can indicate that the ``wall" is not perfect, there may be ``holes" in the wall.}

Since the critical density for the C$^{34}$S(7--6) transition is $\sim 10^7$~\cmc\ \citep[e.g.][]{Codella14}, the gas density in these walls should be at least close to this value. {In principle a noticeable emission can arise at much lower densities \citep{Shirley15} but it requires a high optical depth in the lines and correspondingly high column density. The numerical estimates will be given below.}

{The similarity of the morphology of the CCH and C$^{34}$S emission hints on their common origin.} Observations of dense outflow walls in the CS and CCH lines have been reported recently for both low-mass and high-mass star-forming regions \citep{Bruderer09b, Gomez-Ruiz15, Oya15, Oya18, Leung16,  Zhang18}. The reasons for enhancement of the abundances of these molecules are discussed in these works. Briefly, the CS abundance increases sharply at $ T>100 $~K due to the evaporation of sulfur from dust grains. CCH can be formed in photodissociation regions created by the UV radiation from the central protostar. It can be also formed in the envelope via warm carbon-chain chemistry. Both the high brightness temperatures in the C$^{34}$S and CCH lines, and these chemical models indicate that the temperature in the walls can be rather high, $\ga 100$~K.
The walls can be heated to rather high temperatures by the UV radiation from the massive protostar \citep[e.g.][]{Bruderer09b,Bruderer09a} or by the energy transfer from the high velocity gas \citep[e.g.][]{Hatchell99}.
{Unfortunately the assumption of such high temperature cannot be confirmed by the available observational data.} 
{The radiation temperature in the C$^{34}$S(7--6) line reaches approximately 40~K in the area shown in Fig.~\ref{fig:sio-box}. In case of a high optical depth and the beam filling factor equal to unity this implies the excitation temperature of $ \approx 48 $~K, which places a lower limit on the kinetic temperature.}

We note, however, that the other outflow in this area, originating at the SMA2 core, shows no associated C$^{34}$S or CCH emission.

In this picture the nature of the filamentary structure seen in continuum is not very clear. It is located between the red-shifted and blue-shifted outflow lobes with wide opening angles and may be influenced by both of them. It is not excluded that it is a part of these walls. The CS and CCH relative abundances in this structure are probably lower than in the walls discussed above, where no noticeable continuum emission is detected. The temperature may be lower, too, but not less than the CCH {excitation temperature discussed above}. We assume the kinetic temperature in the continuum filament to be 50~K. Since the preliminary estimates indicate a quite high gas density $\sim 10^8$~\cmc\ \citep{Zin18-raa}, the dust and gas temperatures should be tightly coupled \citep[e.g][]{Banerjee06}, so the dust temperature may be assumed to be equal to the gas temperature. 

The continuum {radiation} temperature along the continuum filament is $\sim 2-3$~K. This implies the optical depth $\sim 0.06$. Assuming the dust opacity {at this wavelength of 1.8~cm$^2$\,g$^{-1}$} according to \cite{Ossenkopf94} we obtain the dust column density $\sim 0.03$~g\,\cms\ and assuming the mean molecular weight per hydrogen molecule of 2.8 \citep{Kauffmann08}, we further estimate the H$_2$ column density $\sim 7\times 10^{23}$~\cms. Assuming the thickness of the filament the same as its width ($\sim 1500$~AU) we obtain the mean hydrogen volume density $n\sim 3\times 10^7$~\cmc. Both the inferred column density and the volume density are lower than the preliminary values mentioned earlier \citep{Zin18-raa} as the adopted dust temperature here is higher. {However the thickness of this structure may be larger} than the observed width and therefore this value for density should be considered as an upper limit. 
The total mass of the filamentary structure seen in continuum is about 35~M$_\odot$ (under the assumption of the dust temperature equal to 50~K). 

Our non-LTE modeling with RADEX \citep{vdTak07} under the assumptions of $T=50$~K, $n = (1-3)\times 10^7$~\cmc\ and $\Delta V = 1.5$~\kms\ for the CS $J=7-6$ transition (C$^{34}$S is absent in the RADEX database) gives an estimate for the C$^{34}$S column density $N(\mathrm{CS}) \sim 5\times 10^{13}$~\cms, which implies the C$^{34}$S relative abundance $X(\mathrm{C^{34}S})\sim 7\times 10^{-11}$. The optical depth in the line is $ \sim 0.3 $. This estimate is consistent with the typical values of the C$^{34}$S abundance in high-mass star-forming cores \citep[e.g.][]{Shirley03} although an order of magnitude lower than our estimate for S255 from single-dish observations \citep{Zin09}. {The fit for $n = 10^6$~\cmc\ gives $N(\mathrm{C^{34}S}) \sim 2\times 10^{14}$~\cms\ and the optical depth in the line $ \sim 1.7 $ (even higher in the lower transitions). This implies a huge optical depth in the lines of the main isotopologue. Taking into account the results of the CCH excitation analysis, we prefer the density estimate of $n = (1-3)\times 10^7$~\cmc, which is consistent with the C$^{34}$S, CCH and continuum data.}

However in the regions of the strong C$^{34}$S emission without continuum counterpart the C$^{34}$S relative abundance should be  much higher {if the dust to gas mass ratio has the ``standard" value}. {Under this assumption}, the hydrogen column density in these regions should be at least several times lower than in the continuum filament, in order to explain the absence of the continuum emission. This implies also a lower value of the mean volume density in these regions. 
{As shown above, the observed C$^{34}$S(7--6) line can be modelled at relatively low densities $n = (1-3)\times 10^6$~\cmc. For example, at $T_{kin}=100$~K and $n = 3\times 10^6$~\cmc\ the required C$^{34}$S column density is $N(\mathrm{C^{34}S}) \sim 2.2\times 10^{14}$~\cms\ with the optical depth in the line $ \sim 1.3 $. This implies the CS relative abundance $\sim 10^{-7}$, which is close to the highest values obtained in the models \citep{Bruderer09b}. 
These requirements can be relaxed by assuming} that the ``wall" structure may have a clumpy nature, with which the absence of strong continuum emission is resulted from the relatively low mean volume density and the C$^{34}$S excitation is achieved in high density clumps. {Fig.~\ref{fig:sio-box} shows that the C$^{34}$S distribution may be rather inhomogeneous on small scales, indeed.  An alternative explanation can be based on the reduced dust to gas mass ratio. The dust grains can be destructed by shocks or swept out somehow. The destruction usually leads to an enhanced SiO abundance, which we do observe in some compact regions (see below). However there is no sign that such process could happen in the rather extended areas of the significant C$^{34}$S emission outside the continuum ``filament". The swept-out dust should be seen somewhere, but we do not see this. Therefore we conclude that this alternative explanation can be rejected.}

\subsection{Outflow from SMA1}

The outflow from SMA1, as observed in CO(3--2), has a wide opening angle. At the same time the compact \Feii\ emission at the ends of the outflow lobes (Fig.~\ref{fig:channel}), which coincides with the radio continuum knots observed with the VLA \citep{Cesaroni18} indicates the presence of strong bow shocks. Probably there is a narrow jet, creating these bow shocks, and a surrounding wind with a wide opening angle. {The orientation of the red-shifted jet is apparently somewhat different from that found by \cite{Howard97} (which is indicated by the dashed line in Fig.~\ref{fig:channel}).} The radio continuum data imply the presence of the ionized gas, which is apparently concentrated in the narrow jet \citep{Zin15,Cesaroni18}. Such picture is consistent with the existing outflow models \citep[e.g.][]{Arce07} but has been observed mainly in low mass protostars \citep{Anglada18}. Probably the best example of such two-wind outflow from a massive protostar is Cep A HW2, where a fast narrow jet and a slower wide-angle outflow have been identified from the water maser observations \citep{Torrelles11}.

In our case the CO emission is relatively weak along the outflow axis and peaks near the outflow walls.
A rather strong SiO emission is observed mainly at the edges of high velocity flows traced in CO, especially in the red-shifted outflow lobe (Fig.~\ref{fig:sio+cont+outflow}). In particular near the position outlined by the dashed box in Fig.~\ref{fig:sio+cont+outflow} it looks as a narrow lane oriented perpendicular to the CO flow (at its end). Most probably it is associated with a shock, as usually expected for SiO. The morphology of the SiO emission is consistent with this assumption \citep[e.g.][]{Masson93}.

Figs.~\ref{fig:sio+cont+outflow} and \ref{fig:sio+c34s-pv} show that the peak of the SiO emission is located at the inner side of the C$^{34}$S wall, although a fainter emission is observed also beyond the wall. 
The SiO and C$^{34}$S velocities are significantly different (Figs.~\ref{fig:sio+c34s-pv}, \ref{fig:sio+c34s-sp}). In principle the SiO and C$^{34}$S emission regions can be spatially separated on the line of sight. However there are signs of interaction between them. While the SiO velocity increases towards the wall on the inner side, it begins to decrease on the other side. The C$^{34}$S velocity dispersion is somewhat larger on the inner side of the wall. 
A significant velocity difference between the wall and the outflowing gas means that a momentum and energy transfer from the outflow to the wall is possible. 

Properties of the high velocity molecular gas can be estimated from the CO(3--2) data. The total velocity range of the CO(3--2) emission is from about --30~\kms\ to about 40~\kms. The systemic velocity of the driving source is about 5~\kms \citep{Zin15}. There are different estimates of the inclination angle in this system but most probably it is rather large, about 80$^\circ$ \citep{Boley13,Cesaroni18}. In this case the maximum outflow velocity is about 200~\kms. {At the same time the characteristic outflow velocity as defined by \cite{Bally83} (see also \citealt{Cabrit90}) is $\sim 58$~\kms\ for the red-shifted lobe and $\sim 50$~\kms\ for the  blue-shifted lobe. Taking the spatial extent of the red-shifted outflow lobe about 8000~AU (as marked by the \Feii/radio knot) we obtain the outflow dynamic time scale about 700~years.} It is worth mentioning that for the SW radio knot \cite{Cesaroni18} found the velocity of $\sim 660$~\kms\ from the variations of its morphology on the time interval of 26 years. However they found no variations for the NE knot and noticed that the ejections can be asymmetric. For the further estimates {we use the obtained estimate of the dynamic time scale of 700 years} for the red-shifted lobe. 

Here we consider local properties of the red-shifted outflow lobe in the interaction zone near the position marked by the yellow circle in Fig.~\ref{fig:sio+cont+outflow}. The brightness temperature in CO is $\sim 100$~K. From the red wing of the CO(3--2) spectrum we obtain the CO column density $N(\mathrm{CO})\ga 1.3\times 10^{18}$~\cms\ under the assumptions of the LTE conditions and low optical depth, which translates into $N(\mathrm{H_2})\ga 1.3\times 10^{22}$~\cms. The thickness of the emission layer seems to be not larger than 2000--3000~AU. Then, the volume density of the high velocity gas is $\ga 3\times 10^5$~\cmc. Such rather high density is consistent with our previous observations of the HCN (4--3), HCO$^+$(4--3) and CS(7--6) emission in this outflow \citep{Zin15}. This estimate shows that the density contrast between the walls and the high velocity gas is not high, probably less than an order of magnitude. 

{The total mass of the high velocity gas estimated from the flux in the CO line wings integrated over the outflow region is $\ga 0.3$~M$_\odot$ (assuming a low optical depth). The upper limit for this estimate is determined by the CO excitation temperature. For example, for $ T_{ex}(\mathrm{CO})=300 $~K the outflow mass estimate is $\sim 0.6$~M$_\odot$ (still assuming a low optical depth).} 
For the outflow {dynamic time scale of 700 years} this estimate gives the mass loss rate $\dot{M}\ga 4\times 10^{-4}$~M$_\odot$\,year$^{-1}$. {\cite{Wang11} estimated the total mass of the outflow from the single-dish observations at 2.9~M$_\odot$. However, this estimate refers to much larger scales, which greatly exceed the field of view of the current observations and does not separate the outflows from the SMA1 and SMA2. The mass loss rate obtained by \cite{Wang11} is the same as the lower limit given above.}

From the comparison of the SiO(8--7) and CO(3--2) intensities at the same velocities, under the assumption of the LTE conditions and low optical depth for both molecules, we obtain the SiO abundance relative to CO $N(\mathrm{SiO})/N(\mathrm{CO})\sim 10^{-4}$. This implies the SiO abundance relative to hydrogen $X(\mathrm{SiO}\sim 10^{-8}$, which represents an enhancement in comparison with a quiescent gas by at least 3 orders of magnitude \citep{Martin-Pintado92}. {This value is close to the highest SiO abundances observed in the outflows in HMSF regions \citep[e.g.][]{Tercero11, Sanchez-Monge13, Hervias-Caimapo19}.} However, the assumption of the LTE conditions is probably not valid for SiO. The deviations from LTE will make the estimate of the SiO abundance even higher.
Our modeling with RADEX \citep{vdTak07} shows that under these conditions the optical depth for both SiO and CO transitions is moderate ($\la 1$) and should not significantly influence the results.

The projected velocity difference between the wall and the high velocity gas reaches $v\approx 10$~\kms. Although the high velocity gas should move mainly along the wall, the extent of the SiO emission perpendicular to the outflow direction implies a significant transverse velocity. This is expected for an adiabatic bow shock \citep{Masson93}, which we probably observe in SiO.

The thickness of the walls seen in C$^{34}$S and CCH is about the same as the width of the continuum filamentary structure, i.e. $ \sim 1000-2000$~AU. The gas velocity dispersion in the walls is low. The C$^{34}$S line width for the spectrum shown in Fig.~\ref{fig:sio+c34s-sp} is approximately 1.2~\kms, which corresponds to the one-dimensional velocity dispersion of about 0.5~\kms. This velocity dispersion is mostly non-thermal, since the thermal dispersion for C$^{34}$S is about 0.13~\kms\ at 100~K. However it is lower than or comparable to the sound speed, which is approximately 0.6~\kms\ at 100~K (for the mean molecular weight per free particle $\mu = 2.33$).

An important question is whether these walls are created by the observed outflowing gas or they have been created earlier and channel the outflow. The general morphology of these structures in comparison with the distribution of the high velocity gas suggests that the walls immediately surrounding the observed outflow are probably created by this outflow or at least strongly influenced by it. At the same time  Fig.~\ref{fig:sio+cont+outflow} makes an impression that in some parts the high velocity gas encounters preexisting dense clumps traced in the C$^{34}$S line. The central source demonstrates signatures of episodic bursts. In Figs.~\ref{fig:channel},\ref{fig:c34s-chmaps} we can see arc-like structures which are not related to the currently observed outflow. They may well be remnants of the previous events. {It is not excluded that in the past there could be wind streams from the SMA1 directed to the north (as we see now in the south). Such streams could trigger star formation at the SMA2 site. Of course, this is just a speculation.}

The next obvious question is how long such walls can survive. Probably they are not gravitationally bound and will disperse in about a crossing time $ t_c \sim L/\sigma_v $, where $ L $ is the thickness and $ \sigma_v $ is the velocity dispersion. Taking $ L\sim 1000 $~AU and $ \sigma_v \sim 0.5$~\kms, as follows from the C$^{34}$S data, we obtain $ t_c \sim 10^4 $ years. The age of the currently observed outflow, estimated from its extension and velocity, is apparently much lower, not more than $ \sim 10^3 $~years. The low velocity dispersion in the walls may be explained by the rapid decay of the supersonic turbulence.

A related question is how long we can see the C$^{34}$S and CCH emission after the influence of the outflow disappear. The answer depends on the cool-down time and the chemical evolution. In case of dust cooling the cool-down time in the optically thin regime can be very short \citep{Banerjee06}. However it requires a tight thermal coupling between gas and dust for efficient gas cooling, which is achieved at $n\ga 10^{7-8}$~\cmc. Since the density in the walls seems to be lower, gas can keep a higher temperature than the dust. Then, we estimated the optical depth of the continuum ``filament" at $\sim 0.05$. For the $\lambda^{-2}$ dependence of the optical depth, this structure will be opaque at the peak of the dust emission ($\sim 50$~$\mu$m for $T_\mathrm{d}\sim 50$~K). The dust column density in the walls is apparently lower but even if it is lower by an order of magnitude, they will be opaque, too. Therefore, the dust cooling may be not very effective here. The cooling rates for molecular cooling calculated by \cite{Neufeld95} imply the cool-down time for our conditions of several thousand years, i.e. comparable to the dispersion time. A typical chemical evolution time is of the same order of magnitude \citep[e.g][]{Burkhardt19}. Therefore, episodic ejections with intervals of $\la 1000$ years could create the observed structure.

\subsection{SMA2}
Our ALMA data show a well-collimated high-velocity extended outflow from SMA2, which is almost parallel (in projection) to the outflow from SMA1. It is not clear what is happening on the larger scales. The previous SMA and IRAM-30m data \citep{Wang11,Zin15} show a single outflow from the S255IR complex at larger distances. It is not excluded that the outflows from SMA1 and SMA2 can merge. 

\begin{figure}
\includegraphics[width=\columnwidth]{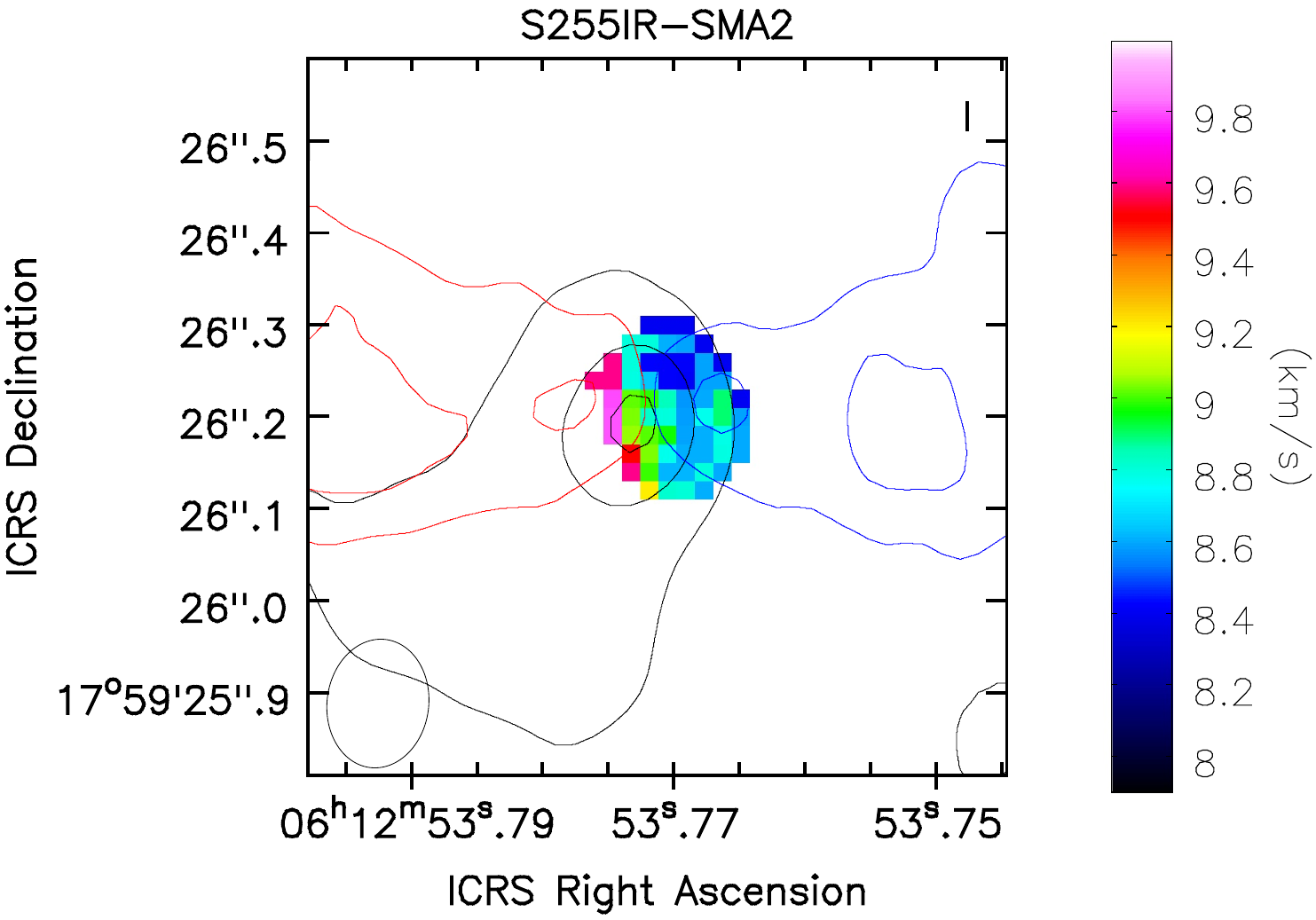} 
\caption{The map of the C$^{34}$S(7--6) 1st moment toward the SMA2 core overlaid with contours of the CO(3--2) high-velocity red-shifted and blue-shifted emission (red and blue contours, respectively) and 0.9~mm continuum emission (black). {The contour levels are at 8, 12, 16~mJy\,beam$^{-1}$ for continuum, at 0.3, 0.6~Jy\,beam$^{-1}$\,\kms\ for the red-shifted CO emission and at 0.17, 0.34~Jy\,beam$^{-1}$\,\kms\ for the blue-shifted CO emission.} The ALMA beam {for C$^{34}$S(7--6)} is shown in the lower left corner.}
\label{fig:sma2-mom1+outflow}
\end{figure}

\begin{figure}
\includegraphics[width=\columnwidth]{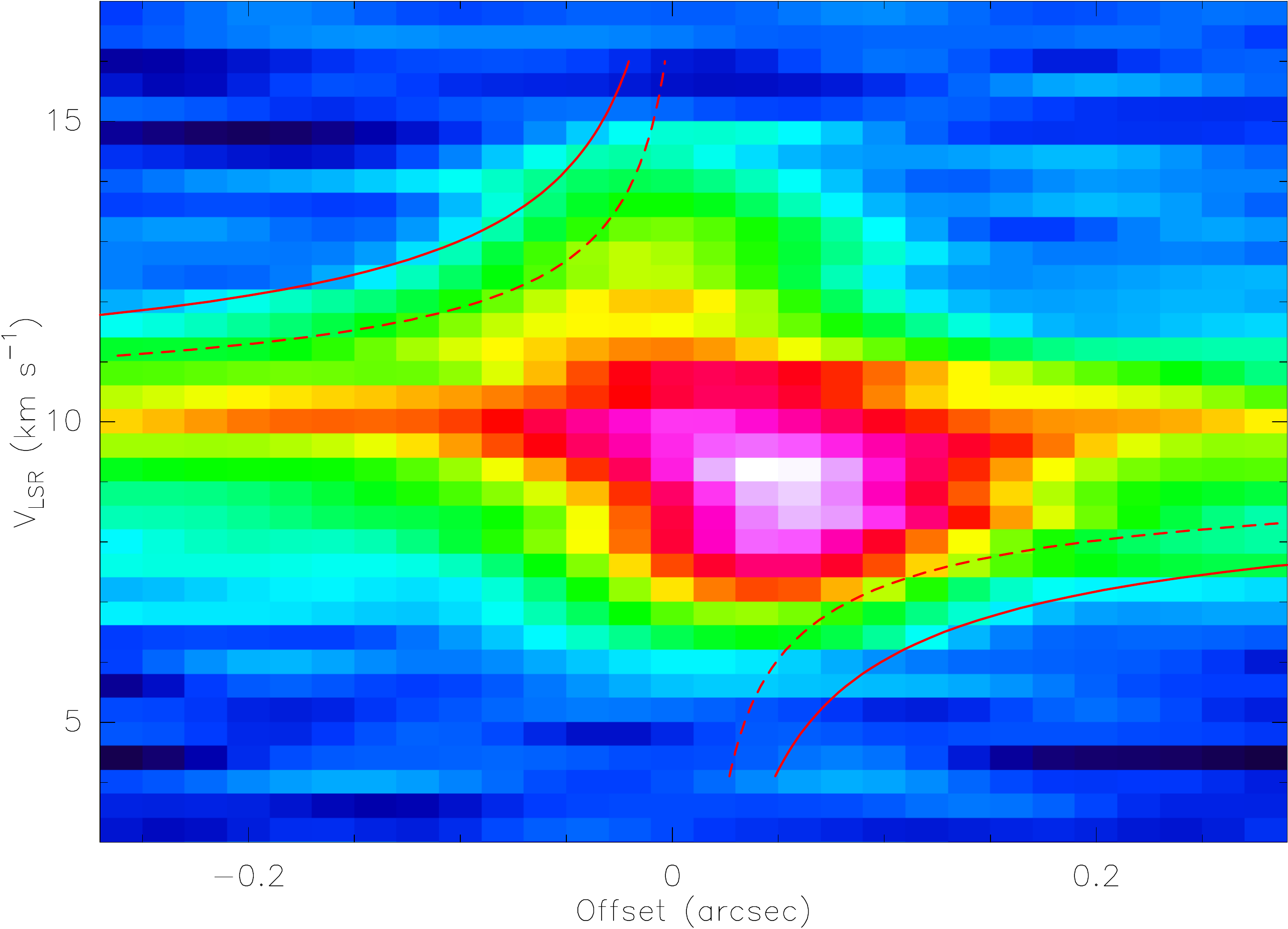} 
\caption{The position-velocity diagram in the C$^{34}$S(7--6) line along the major axis of the SMA2 core (PA = 150$^\circ$). The curves correspond to Keplerian rotation around the central mass of $M \sin^2 i = 2.2$~M$_\odot$ (solid) and $M \sin^2 i = 1.5$~M$_\odot$ (dashed).}
\label{fig:sma2-c34s-pv}
\end{figure}

The close-up view of the SMA2 region is shown in Fig.~\ref{fig:sma2-mom1+outflow}. Here the map of the 1st moment of the C$^{34}$S(7--6) emission is overlaid with the contours of the CO(3--2) outflow and 0.9~mm continuum emission. In C$^{34}$S we see a compact core with {an apparent} velocity gradient. Both the continuum and C$^{34}$S(7--6) integrated intensity maps show an elongated structure with the position angle of the major axis about $150^\circ - 160^\circ$. {However the position angle of the velocity gradient seems to be different.} The size in continuum is $\sim 0\farcs37\times 0\farcs25$ ($\sim 670\times 450$~AU). The flux density in continuum is about 60~mJy. The kinematics of this core is illustrated in more detail in Fig.~\ref{fig:sma2-c34s-pv}, where we present the position-velocity diagram for this core along the major axis in the C$^{34}$S(7--6) line.

The data {are consistent with the assumption of} a Keplerian-like rotation of the core. Apparently it resembles a disk, {although the achieved spatial resolution is not sufficient for a firm conclusion}. The continuum flux gives an estimate of the disk mass $\sim 0.4$~M$_\odot$ (assuming the dust temperature of 50~K and a normal gas-to-dust mass ratio of 100). For the measured size of SMA2 this mass implies the mean gas density $n\sim 3\times 10^8$~\cmc. At the same time the dynamical central mass obtained from the rotation velocity is much higher, $M \sin^2 i \sim 2$~M$_\odot$, where $i$ is the inclination angle (Fig.~\ref{fig:sma2-c34s-pv}).

The observed aspect ratio for this core indicates a substantial inclination of the probable disk. However, it hardly can be as large as assumed for the SMA1 core ($\sim 80^\circ$) because in this case the {terminal} velocity of the molecular outflow would reach $\sim 400$~\kms\ {and the characteristic velocity $\sim 300$~\kms}, which looks unrealistic. {For the inclination angle of $\sim 70^\circ$ these values will be approximately 2 times lower and close to the highest reported values for the outflows \citep[e.g.][]{Bachiller90}. From Fig.~\ref{fig:co-outflow} it is evident that the peak outflow velocity increases with the distance from the driving source, which is rather usual for molecular outflows \citep{Arce07}.}

\begin{figure}
\includegraphics[width=\columnwidth]{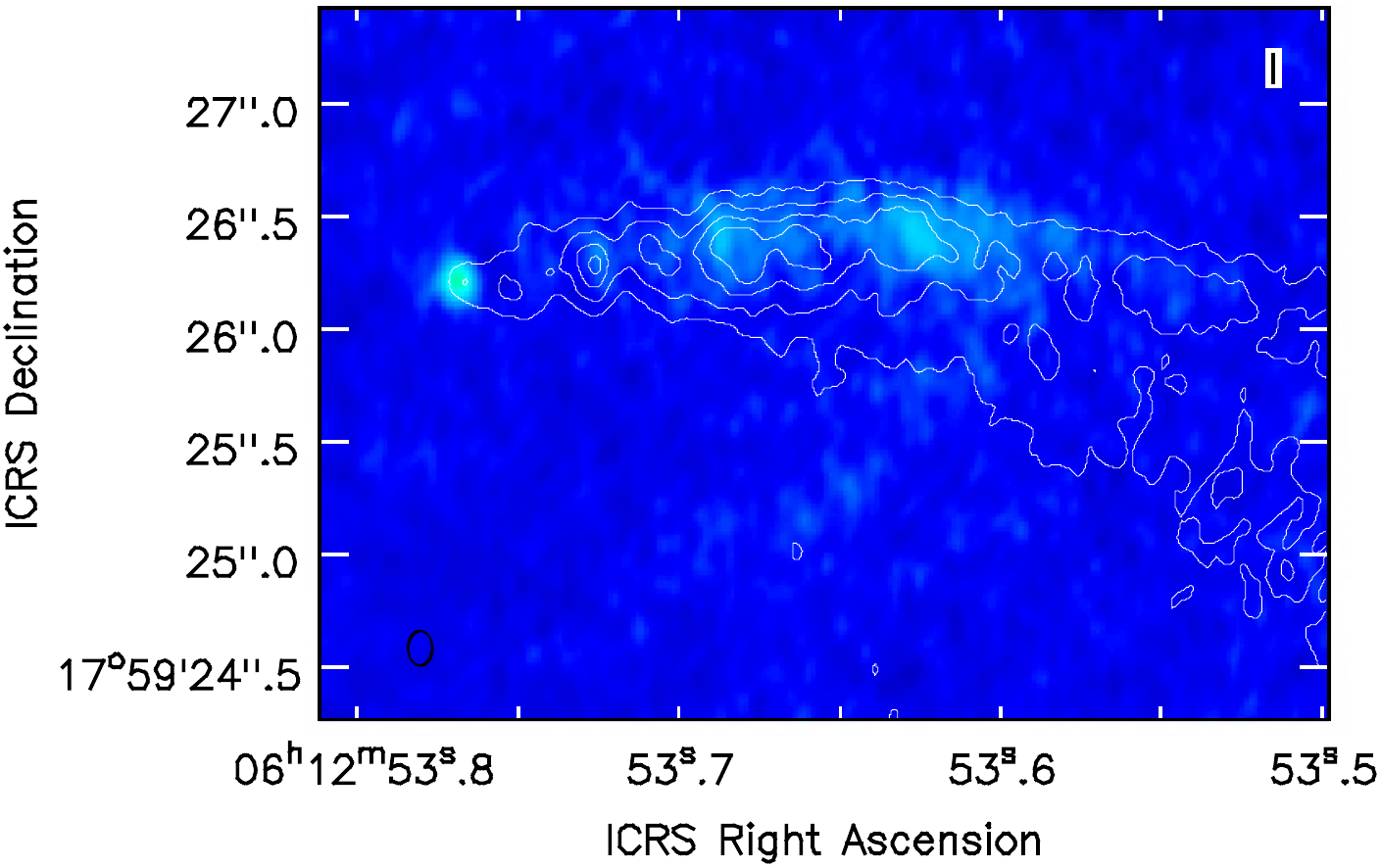} 
\caption{The image of the blue-shifted outflow lobe from SMA2 in the SiO(8--7) line emission. {The velocity range is from --42 to --9~\kms.} The contours show the high velocity {blue-shifted} CO emission. The contour levels are from 0.2 to 1.0 in step of 0.2~~Jy\,beam$^{-1}$\,\kms.}
\label{fig:sma2-sio+co}
\end{figure}

We see a rather strong SiO(8--7) emission in the blue-shifted outflow lobe, which correlates well with the CO emission (Fig.~\ref{fig:sma2-sio+co}). The enhancement of the SiO abundance here, as indicated by the SiO(8--7)/CO(3--2) intensity ratio, is even higher (by a factor of $\sim 3$) than in the shocked region in the outflow from SMA1. However, no noticeable SiO emission is observed from the red-shifted lobe. Distribution of the C$^{34}$S(7--6) emission (Fig.~\ref{fig:channel}) hints that the blue-shifted lobe probably propagates in a denser medium, containing more dust. This can explain a higher SiO abundance in this lobe due to the dust grain destruction.

{The high collimation of the outflow and the Hubble-type velocity-distance relation imply a jet-driven bow shock model for the outflow \citep{Arce07}.}
{The outflow is apparently bent (Figs.~\ref{fig:channel},\ref{fig:sma2-sio+co}). Such bending can be interpreted as a result of disk precession induced by misalignment between the disk and the orbital plane in a binary system {taking into account that jets are always launched perpendicular to the disk plane} \citep{Monin07}.} 

Then, it is easy to see that the outflow from SMA2 consists of a chain of knots separated by about 0.2--0.3 arcsec, i.e. 350--500~AU. Most probably this indicates periodic ejections from the central object. {The velocities of the knots closest to the driving source are apparently significantly lower than the terminal and characteristic velocities mentioned above (Fig.~\ref{fig:co-outflow}). They imply the velocities in the plane of the sky $\sim 30$~\kms, which gives} the time interval between these ejections $\sim 50-80$~years. {This can be related to some periodic process in this system, perhaps to the orbital period.}

In general, in SMA2 we {may} witness the formation of a binary system consisting of low or intermediate mass stars.

\subsection{SMA3}
The SMA3 core is associated with the NIRS1 infrared source \citep{Tamura91}. The measured flux in continuum at 0.9~mm implies the mass of this core $\sim 0.3$~M$_\odot$ (under the previous assumptions about the dust properties and assuming the dust temperature of 50~K). The deconvolved size of the continuum source is approximately 0.1 arcsec (Table~\ref{table:cont-mm}), i.e. about 180~AU and no elongation is seen. The corresponding mean gas density is very high, $n\sim 10^{10}$~\cmc. {However the dust temperature in vicinity of this luminous IR source can be higher than assumed above. Then the estimates of the mass and density will be correspondingly lower.} No clear velocity gradient is observed in the C$^{34}$S(7--6) line. In general the source looks isolated and not related to the filamentary continuum structure discussed above. 

The NIR polarization measurements show convincing evidences for a bent {``S-shape"} outflow from this object oriented approximately in the south-north direction \citep{Simpson09}. {On this basis \cite{Simpson09} suggested that this can be a part of a binary.} However our data do not indicate any molecular outflow from this core. A reasonable assumption is that the outflow lies in the plane of the sky, so that we do not see any high velocity emission and at velocities close to the systemic one the morphology of the CO emission is too complicated.

\subsection{New cores}
The newly detected cores labelled here as SMM1 and SMM2 (Fig.~\ref{fig:filament}, Table~\ref{table:cont-mm}) have flux densities at 0.9~mm of 15 and 10~mJy, respectively. These values correspond to masses of $\sim 0.3$~M$_\odot$ and $\sim 0.2$~M$_\odot$, respectively (assuming the dust temperature of 20~K). The mean gas density for these cores is $n\sim 3\times 10^9$~\cmc. {We see no molecular emission clearly associated with these cores. It means that the molecules may be frozen on the dust grains. The absence of the line emission does not allow an estimation of the virial mass but the appearance of the cores suggests that they are probably gravitationally bound. At the same time they have no embedded IR sources.} They may represent {low-mass} prestellar cores {according to the widely adopted classification \citep{Andre00,diFrancesco07}}.

\section{Conclusions}
The main results of this study are the following.

1. Our data show a wide-angle molecular outflow from the massive YSO S255 NIRS3, embedded in the dense core S255IR-SMA1. {The outflow dynamic time scale is $ \sim 700 $ years.} This wide-angle outflow surrounds the narrow jet traced at the IR wavelengths and in radio continuum. There are bow shocks at the ends of the jet observed in the radio continuum and in the \Feii\ emission. The CO emission is weak along the outflow axis and peaks near the outflow walls. There are other shocks at the ends of the CO high velocity flows, observed in the SiO(8--7) line. The SiO abundance there is enhanced by at least 3 orders of magnitude. The density of the high velocity gas is $\ga 4\times 10^5$~\cmc.

2. We detected warm and dense walls around this wide-angle molecular outflow. The walls are observed in the C$^{34}$S(7--6) and CCH $N=4-3$ lines. The gas temperature in the walls is at least $\sim 50$~K as indicated by the line brightness temperature, {perhaps} $\ga 100$~K, as follows from the apparently high CS abundance. The thickness of the walls is $ \sim 1000-2000$~AU. The walls are non-uniform. There are signs of the interaction between the high velocity outflowing gas and the walls. 


In addition to the walls surrounding the observed outflow, there are similar structures not associated with the currently observed outflow. They may represent remnants of the previous ejection events. 

3. In continuum at 0.9 mm we see a very narrow ($\sim 1000-1800$~AU) and dense ($n\sim 3\times 10^7$~\cmc\ assuming the cylindrical geometry) filamentary structure {with} at least two velocity components. The mass estimated from the continuum emission is about 35~M$_\odot$. The SMA1 and SMA2 cores are apparently associated with this structure, while the SMA3 core looks isolated. The nature of this continuum structure is not quite clear. It may be influenced by both red-shifted and blue-shifted outflow lobes and it is not excluded that it is a part of the outflow walls.

4. The CO(3--2) data show a collimated and extended high velocity outflow from another dense core in this area, SMA2. The outflow direction {in projection} is approximately parallel to the outflow from SMA1. The SMA2 core is elongated and rotating in accordance with the Keplerian law. The central mass is $M  \ga 2$~M$_\odot$, while the mass of the probable disk is $\sim 0.4$~M$_\odot$. 
We see a rather strong SiO(8--7) emission in the blue-shifted outflow lobe. The outflow from SMA2 consists of a chain of knots separated by about 0.2--0.3 arcsec, i.e. 350--500~AU. Most probably this indicates periodic ejections from the central object with the period of 30--50 years. 

5. The SMA3 core is associated with the NIRS1 infrared source. The measured flux in continuum at 0.9~mm implies the mass of this core $\sim 0.3$~M$_\odot$ {(assuming the dust temperature of 50~K)}. 
Although the NIR polarization measurements show convincing evidences for an outflow from this object, our data do not indicate any molecular outflow from this core. A reasonable assumption is that the outflow lies in the plane of the sky.

6. We detected two new compact cores in this area with masses $\sim 0.2-0.3$~M$_\odot$ and sizes $\sim 200-300$~AU. Their mean gas densities are $n\sim 3\times 10^9$~\cmc. They may represent low-mass prestellar cores.

\begin{acknowledgements}
We are grateful to the anonymous referee for the detailed and constructive comments, which helped to improve the paper.
This research was supported by the Russian Science Foundation (grant No. 17-12-01256) in the part of the data analysis and by the Russian Foundation for Basic Research (grant 18-02-00660) at the data reduction.  S.-Y. Liu and Y.-N. Su acknowledge the support by the Minister of Science and Technology of Taiwan (MOST 108-2112-M-001-048).
This paper makes use of the following ALMA data: ADS/JAO.ALMA\#2015.1.00500.S. ALMA is a partnership of ESO (representing its member states), NSF (USA) and NINS (Japan), together with NRC (Canada), MOST and ASIAA (Taiwan), and KASI (Republic of Korea), in cooperation with the Republic of Chile. The Joint ALMA Observatory is operated by ESO, AUI/NRAO and NAOJ.
\end{acknowledgements}

\bibliographystyle{aasjournal} 
\bibliography{filament}



\end{document}